\pdfoutput=1
\documentclass[preprint]{aastex631}
\usepackage{calc}
\usepackage{CJKutf8}
\usepackage[useregional]{datetime2}

\accepted{on 2025-12-01 for publication in ApJ}

\def\affnssc{State Key Laboratory of Solar Activity and Space Weather,
    National Space Science Center, Chinese Academy of Sciences, Beijing 100190, China;
    \href{mailto:huhd@nssc.ac.cn}{huhd@nssc.ac.cn}}
\def\affucas{University of Chinese Academy of Sciences, Beijing 100049, China}
\def\affbjseu{Space Engineering University, Beijing 101416, China}
\def\affcma{Key Laboratory of Space Weather, National Satellite Meteorological Center,
    China Meteorological Administration, Beijing 100081, China}
\def\affhusb{School of Microelectronics and Physics, Hunan University of Technology and Business, Changsha 410205, China;
    \href{mailto:chenc@hutb.edu.cn}{chenc@hutb.edu.cn}}

\begin{document}

\title{Lateral Deformation of Large-scale Coronal Mass Ejections
  during the Transition from Non-radial to Radial Propagation}
\shorttitle{Lateral Deformation of CMEs in Non-radial to Radial Transition}
\shortauthors{Hu et al.}
\author[0000-0001-8188-9013]{Huidong Hu \texorpdfstring{\begin{CJK*}{UTF8}{gbsn}(胡会东)\end{CJK*}}{}}
\affiliation{\affnssc}
\author[0000-0002-2316-0870]{Chong Chen \texorpdfstring{\begin{CJK*}{UTF8}{gbsn}(陈冲)\end{CJK*}}{}}
\affiliation{\affhusb}
\author[0009-0004-4832-0895]{Yiming Jiao \texorpdfstring{\begin{CJK*}{UTF8}{gbsn}(焦怡明)\end{CJK*}}{}}
\affiliation{\affnssc}
\affiliation{\affucas}
\author[0000-0001-6306-3365]{Bei Zhu \texorpdfstring{\begin{CJK*}{UTF8}{gbsn}(朱蓓)\end{CJK*}}{}}
\affiliation{\affbjseu}
\author[0000-0001-5205-1713]{Rui Wang \texorpdfstring{\begin{CJK*}{UTF8}{gbsn}(王瑞)\end{CJK*}}{}}
\affiliation{\affnssc}
\affiliation{\affucas}
\author[0000-0002-4016-5710]{Xiaowei Zhao \texorpdfstring{\begin{CJK*}{UTF8}{gbsn}(赵晓威)\end{CJK*}}{}}
\affiliation{\affcma}
\author[0000-0003-4716-2958]{Liping Yang \texorpdfstring{\begin{CJK*}{UTF8}{gbsn}(杨利平)\end{CJK*}}{}}
\affiliation{\affnssc}
\affiliation{\affucas}

\begin{abstract}
Many coronal mass ejections (CMEs) initially propagate non-radially,
and then transition to radial propagation in the corona.
This directional transition is a significant process
that determines a CME's space weather effects but remains poorly understood.
Based on multi-wavelength observations,
we investigate the transition from non-radial to radial propagation in the low corona
for two large-scale CMEs from the same active region on the solar limb.
In the beginning, both CMEs move in a non-radial direction,
beneath a system of overlying loops that are roughly parallel to the flux-rope axis.
The CMEs laterally deform by bulging their upper flanks
{in the non-radial stage} toward the higher corona,
which results in the transition to a radial propagation direction
approximately 25\textdegree{} away from the eruption site.
After the directional transition, the {non-radial-stage} upper flank
becomes the leading edge in the radial stage.
Although the overlying loops do not strap over the flux rope,
their strong magnetic tension force constrains the radial expansion of {part of} the CME during the transition
{by acting on the flux-rope legs}.
{A major portion of the} filament is displaced to the southern part of a CME in the radial stage,
which {implies the complexity of observational CME features}.
This study presents the first investigation of the lateral deformation
during the transition of CMEs from non-radial to radial in the low corona,
and makes an essential contribution to the complete CME evolution picture.
\end{abstract}

\section{Introduction}\label{intro}
Solar coronal mass ejections (CMEs) are large-scale expulsions of plasma and magnetic fields
caused by intense eruptions in the solar atmosphere.
Their physical properties vary with the phases of a solar cycle,
where the speed ranges from 20 km s$^{-1}$ to 3000 km s$^{-1}$ near the Sun,
the mass is in the range of $10^{8}$ kg to $10^{14}$ kg,
and the kinetic energy is on the order of $10^{18}$ to $10^{26}$ Joules
\citep{VourlidasHE2010ApJ,WebbH2012LRSP,LiuHZ2017ApJ,AlobaidAW2023ApJ,GandhiPP2024SpWea}.
When a CME encounters the Earth,
it can disrupt the Earth's magnetic fields and inject a large amount of energy and energetic particles into the magnetosphere,
causing a geomagnetic storm, which is a major aspect of space weather.
A severe space weather event can significantly impact the operation of technological infrastructures
and even pose a threat to human health \citep{Howard2014,RileyBL2018SSRv}.
The initiation and evolution of CMEs are major topics in space weather and solar physics research.
A CME may undergo various physical processes during its propagation from the Sun to interplanetary space.
These processes include acceleration and deceleration
\citep[e.g.,][]{SheeleyWW1999JGR,GopalswamyLL2000GeoRL},
collision with other CMEs
\citep[e.g.,][]{LugazFD2012ApJ,LiuHZ2024ApJ},
interaction with the ambient solar wind
\citep[e.g.,][]{SavaniOR2010ApJ,LiuHR2015ApJ,ChenLZ2024ApJ},
and rotation about its propagation direction
\citep[e.g.,][]{VourlidasCN2011ApJ,ThompsonKT2012SoPh,ChenLW2019ApJ}.
These processes can reshape CMEs and alter their space weather effects
\citep[e.g.,][]{LiuLK2014NatCo,LugazTW2017SoPh,RileyBL2018SSRv}.

The propagation direction of a CME is a significant factor in determining its space weather effects
because it determines whether and which part of the CME impacts the Earth.
However, a significant fraction of CMEs are deflected and propagate in a non-radial direction
in the corona or interplanetary space, away from their original direction
\citep[e.g.,][]{MacQueenHC1986JGR,WangSW2004SoPh}.
A CME may be deflected by interacting with another CME
\citep[e.g.,][]{GopalswamyYK2001ApJ,LugazFD2012ApJ}.
For most cases, the deflection in the corona is caused by the imbalance of the background magnetic pressure
\citep[e.g.,][]{ShenWG2011SoPh,KayOE2015ApJ}.
The magnetic pressure imbalance can arise from a coronal hole or an active region on one side of a CME,
where the magnetic field is stronger
\citep[e.g.,][]{GopalswamyMX2009JGRA,PanasencoMV2013SoPh,HuLW2017ApJ,CecereSC2020AdSpR}.
A heliospheric current sheet (HCS) with lower magnetic pressure than the surrounding regions
can also cause a CME to change its direction and propagate toward the HCS
\citep[e.g.,][]{ShenWG2011SoPh,LiewerPV2015SoPh}.
Magnetic reconnection between a CME and its ambient magnetic fields
can also lead to pressure imbalance,
which results in the redirection of the CME
\citep{ZuccarelloBJ2012ApJ}.
Additionally, a CME can move through a non-radial channel that is formed by open magnetic fields near its source region
\citep[e.g.,][]{MoestlRF2015NatCo,WangLD2015ApJ,SahadeVM2025ApJ}.
Some CMEs also erupt non-radially due to an asymmetric magnetic configuration
\citep[e.g.,][]{SunHL2012ApJ,PanasencoMV2013SoPh},
although such asymmetry may suppress CME eruption \citep{LiuJF2024MNRAS}.

Taking different types of non-radial CMEs into account,
the non-radial propagation in the corona is not a rare phenomenon
\citep[see][and references therein]{McCauleySS2015SoPh,ZhangCL2022SoPh,MichalekGY2023ApJ,ZhangOH2024ApJ},
and this process usually completes in the corona below 10 R$_\odot$
\citep[e.g.,][]{ShenWG2011SoPh,GuiSW2011SoPh,KayOE2015ApJ}.
{Non-radial CMEs may become radial upon entering a region
where the drivers of non-radial propagation diminish.
Actually, after the non-radial propagation}, most of these CMEs eventually leave the Sun in a nearly radial direction
that is often about a dozen degrees away from the source positions
\citep[e.g.,][]{GuiSW2011SoPh,KayOE2015ApJ,HuLW2017ApJ}.
The magnetic fields of an erupting flux rope are rooted in the photosphere,
and thus can scarcely move over such a large distance.
The CME must reshape its structure to complete the transition from non-radial to radial,
as it cannot turn its propagation direction like a rigid body.
This directional transition in the corona is an important stage
in the evolution of non-radial CMEs,
yet it lacks thorough investigation and remains poorly understood.

In this article, we investigate the lateral deformation of two large-scale CMEs from the same active region,
both of which erupt in a non-radial direction beneath a system of overlying loops
and eventually transition to radial propagation.
{Non-radially propagating} CMEs are frequently observed,
but how they transition to radial propagation in the corona has not been previously explored.
This study presents the first detailed analysis of the transition
from non-radial to radial propagation of CMEs in the low corona.
This work sheds light on the roles of both the lateral deformation
and the overlying loops in the directional transition of such CMEs,
and provides implications for a comprehensive understanding of CME evolution.

\section{Lateral Deformation and Directional Transition}\label{sec:transition}
\subsection{CME on 2023 March 3}\label{sec:cme03}
\begin{figure*}[ht!]
  \centering
  \includegraphics[width=0.9\textwidth]{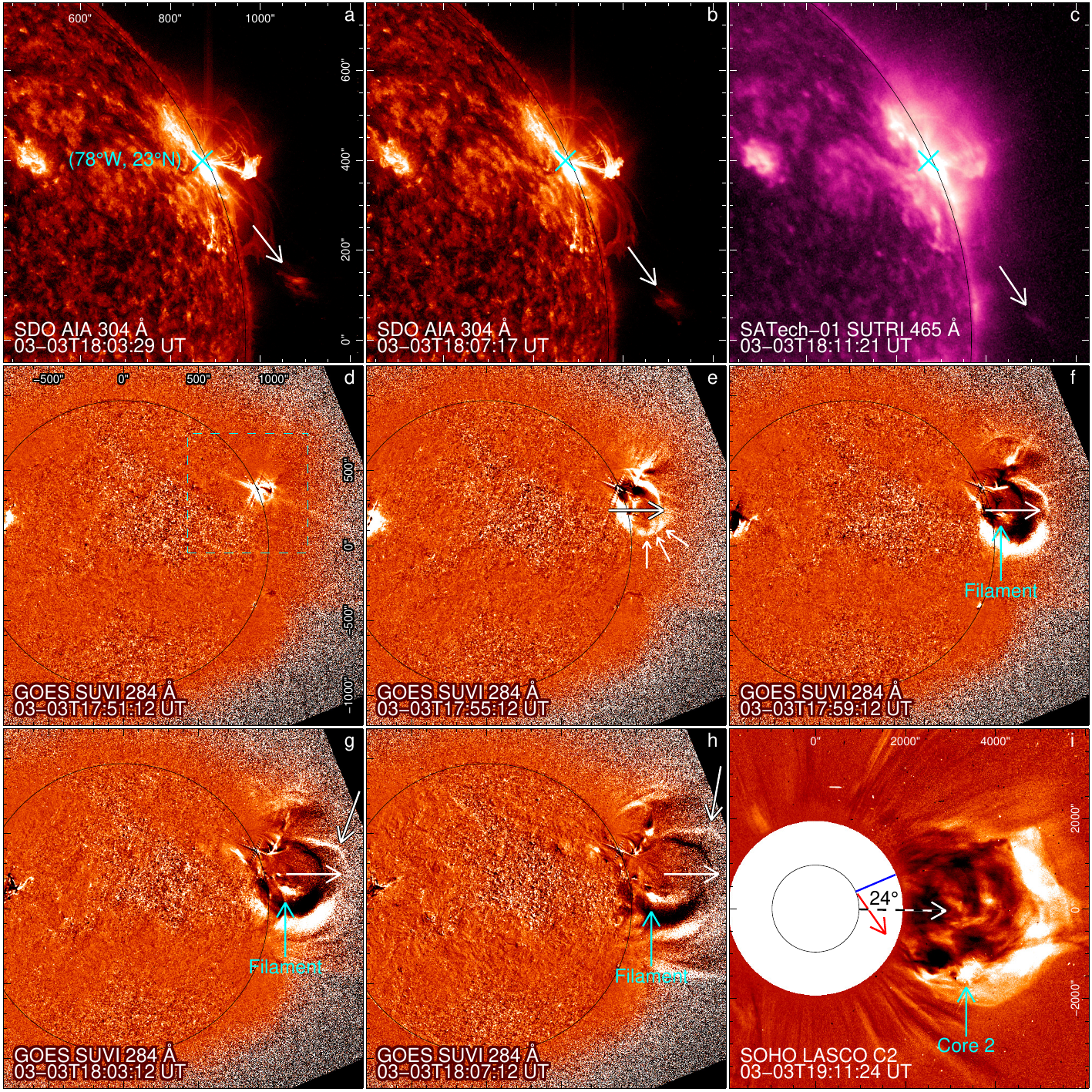}
  \caption{\label{plot03}The CME on 2023 March 3 {observed near the Earth}.
  (a)--(b) The erupted filament observed by SDO AIA 304 \AA, and
  (c) by SATech-01 SUTRI 465 \AA.
  In (a)--(c) the symbol ``\textbf{\texttimes}'' marks the eruption site,
  and the white arrows indicate the erupted filament material
  and its approximate motion direction.
  (d)--(h) The CME in GOES SUVI 284 \AA{} running-difference images,
  {where the three short arrows in (e) mark the CME leading edge in the non-radial stage,
  the rightward arrows in (e)--(h) indicate the bulging upper CME flank,}
  the cyan arrows in (f)--(h) point to the {filament},
  {and the downward arrows in (g)--(h) mark the northward expanding part.}
  (i) The CME in a SOHO LASCO C2 running-difference image,
  where the red arrow indicates the approximate non-radial motion direction of the filament,
  the dashed arrow points to the radial propagation direction of the CME
  (along $\sim$$-$1\textdegree{} latitude,
  which is $\sim$24\textdegree{} away from the eruption site marked by the blue line),
  and the cyan arrow denotes a CME core
  {(``Core 2'', as labeled in Figure \ref{fig:plot03sta}(i))}.
  The dashed square in (d) represents the field of view of (a)--(c).
  For all panels, the abscissa and ordinate are scaled in arcseconds in the helioprojective coordinates.
  {An animation for this figure is available, which contains two parts.
  The first part is for (a)--(b),
  which begins at 2023-03-03T17:49 UT and ends at 2023-03-03T18:20 UT.
  The second part corresponds to (d)--(i),
  where the GOES SUVI images are overlaid on the LASCO C2 images.
  The second part runs from 2023-03-03T17:47 UT to 2023-03-03T19:11 UT,
  with an arrow highlighting the CME leading edge in the LASCO C2 image at 18:11 UT.
  The real-time animation duration is $\sim$8 s.
  In the first part of the animation, the time, 0\textdegree{} latitude, and the solar limb are marked,
  and in the second part, the time and solar limb are shown. } }
  \end{figure*}

The two large-scale CMEs erupted from the same active region (NOAA AR 13234)
near the solar limb on 2023 March 3 and 4, respectively.
The first CME is associated with an X2.1 class flare that peaked at 17:52 UT on March 3 \citep{LiJS2024ApJ}
and is located around (78{\textdegree}W, 23{\textdegree}N).
Figure \ref{plot03}(a)--(c) show
the erupted filament observed at 304 \AA{} by the Atmospheric Imaging Assembly
onboard the Solar Dynamics Observatory \citep[SDO AIA,][]{LemenTA2012SoPh}
and at 465 \AA{} by the Solar Upper Transition Region Imager
onboard the Space Advanced Technology demonstration satellite \citep[SATech-01 SUTRI,][]{BaiTD2023RAA}.
The characteristic temperatures for the two extreme-ultraviolet (EUV) wavelengths
of 304 \AA{} and 465 \AA{} are $\sim$0.5 million Kelvin.
These observations show that the filament material is moving away from the eruption site
in a non-radial, nearly horizontal direction with respect to the solar surface.
The CME structure is captured at 284 \AA{} by the Solar Ultraviolet Imager \citep[SUVI,][]{DarnelSB2022SpWea}
onboard the Geostationary Operational Environmental Satellite (GOES) 16.
The SUVI has a larger field of view (FOV; 1.6 R$_\odot$) than AIA and SUTRI,
and its level-2 data are used in this study,
which have a cadence of 4 minutes with a high dynamic range \citep{DarnelSB2022SpWea}.
As shown in Figure \ref{plot03}(d)--(f),
the CME is expanding, and its leading edge
(indicated by the three white arrows in Figure \ref{plot03}(e)) is moving in a non-radial direction.
Around 18:03 UT, the upper flank of the CME in the non-radial stage {(i.e., the outermost edge)}
is bulging toward the higher corona,
which is marked by the {rightward white arrows in Figure \ref{plot03}(e)--(h).
During the bulging, a brief northward expansion is observed
at the location marked by the downward arrows in Figure \ref{plot03}(g)--(h)}.

After leaving the FOV of GOES SUVI, the CME is imaged by the Large Angle and Spectrometric Coronagraph
onboard the Solar and Heliospheric Observatory \citep[SOHO LASCO;][]{DomingoFP1995SoPh}.
The propagation direction and the morphology of the CME in the FOV of SOHO LASCO C2
are estimated by the Graduated Cylindrical Shell (GCS) fitting method \citep{ThernisienHV2006ApJ}.
The process and parameters of the GCS fitting are given in Appendix \ref{app:gcs}.
{The GCS fitting is based on the coronagraph images of SOHO and the Solar Terrestrial Relations Observatory Ahead
\citep[STEREO A,][]{KaiserKD2008SSRv},
while the two spacecraft are separated by only about 12\textdegree{} and
observe both CMEs as limb events as if from a single viewpoint.}
Based on single-viewpoint observations,
the GCS model {can obtain} the latitude and height
by fixing the other four parameters for limb CMEs
\citep[e.g.,][]{ThernisienHV2006ApJ,ZhaoLH2019ApJ}.
The GCS result shows that {the CME propagation direction in LASCO C2
is approximately radial and gradually changes from 9\textdegree{} to a final $-$1\textdegree{} in latitude.
This final propagation direction} is $\sim$24\textdegree{} away from the eruption site,
as shown in Figure \ref{plot03}(i).

\begin{figure*}[ht!]
  \centering
  \includegraphics[width=0.9\textwidth]{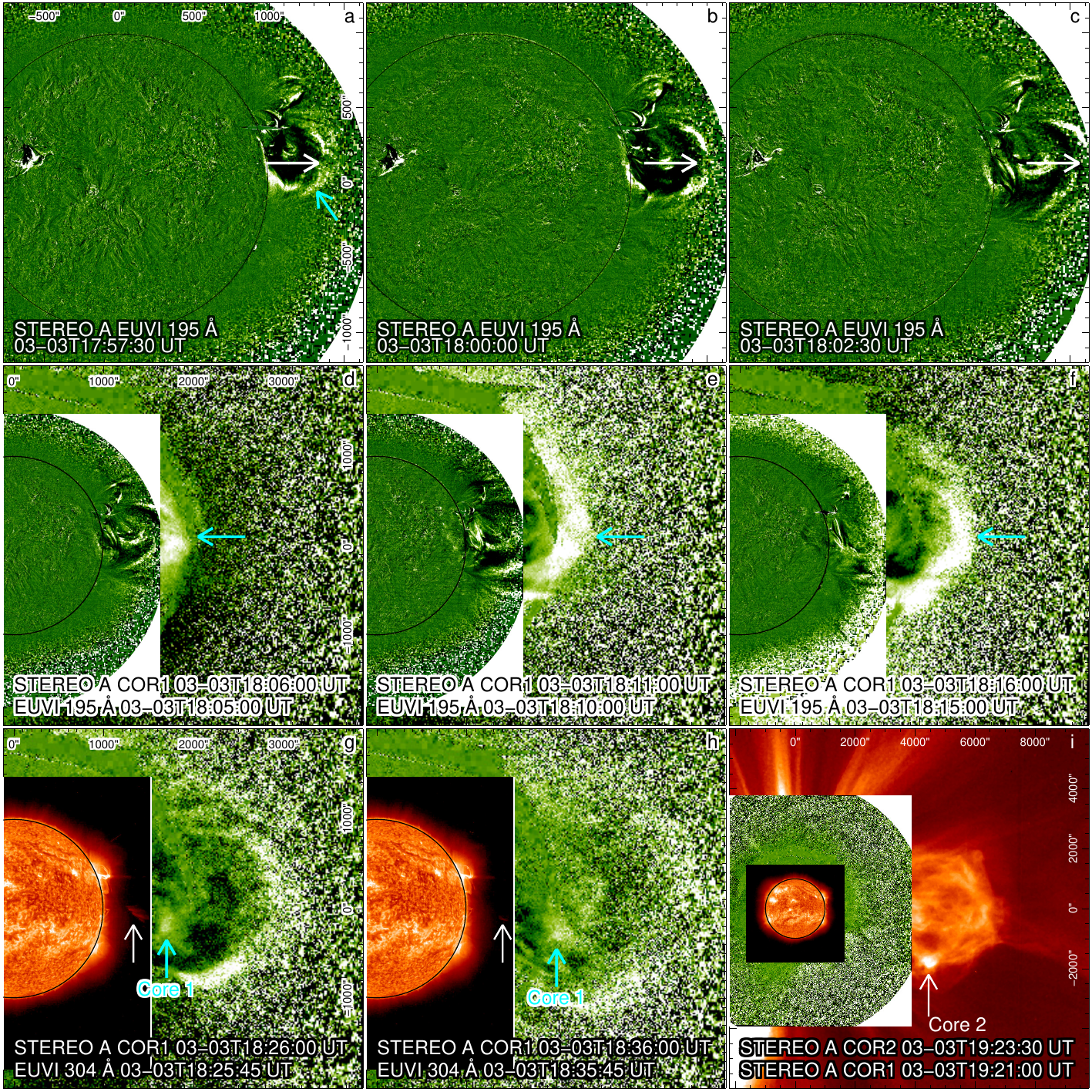}
  \caption{\label{fig:plot03sta}{The CME on 2023 March 3 observed by STEREO A.
  (a)--(c) Running-difference images of STEREO A EUVI 195 \AA{},
  where the rightward white arrows mark the bulging upper flank,
  and the cyan arrow in (a) indicates the CME leading edge in the non-radial stage.
  (d)--(f) Running-difference images of STEREO A COR1 overlaid with running-difference images of EUVI 195 \AA{},
  where the cyan arrows denote the bulged flank entering COR1 as the leading edge in the radial stage.
  (g)--(h) Running-difference COR1 images combined with EUVI 304 \AA{} images,
  where the white arrows mark {the trailing part} of the erupted filament in EUVI 304 \AA{},
  and the cyan arrows indicate {a CME core (``Core 1'') in COR1 corresponding to the leading part of the filament}.
  (i) STEREO A COR2 image superimposed with a running-difference COR1 image and an EUVI 304 \AA{} image,
  where the arrow points to {another CME core
  (``Core 2'', also shown in Figure \ref{plot03}(i))}.
  The axes are labeled similarly to Figure \ref{plot03}.
  An animation for this figure is available,
  which has the same field of view as (i),
  where the COR2 images are overlaid with running-difference images of COR1 and EUVI 195 \AA{}.
  The animation begins at 2023-03-03T17:45 UT and ends at 2023-03-03T19:57 UT,
  and the real-time duration is $\sim$3 s.
  The times are displayed in the animation.}
  }\end{figure*}

Figure \ref{fig:plot03sta} presents the observations of the CME on 2023 March 3
from the Sun-Earth Connection Coronal and Heliospheric Investigation
\citep[SECCHI,][]{HowardMV2008SSRv} onboard STEREO A.
The Extreme Ultraviolet Imager (EUVI), COR1, and COR2 of SECCHI
have FOV outer limits of 1.7 R$_\odot$, 4 R$_\odot$, and 15 R$_\odot$, respectively,
which can continuously track a CME in the corona.
STEREO A was roughly 12\textdegree{} east of the Earth during the events,
and observed the March 3 CME as a limb eruption at $\sim$90\textdegree{} west longitude from its perspective.
The EUVI 195 \AA{} running-difference image in Figure \ref{fig:plot03sta}(a)
shows that the CME is approximately symmetrical about the non-radial propagation direction in the beginning,
where the cyan arrow indicates the CME leading edge in the non-radial stage
(similar to the three small white arrows in Figure \ref{plot03}(e)).
Then the upper CME flank bulges, as marked by the rightward white arrows in Figure \ref{fig:plot03sta}(b)--(c),
which is also consistent with the GOES SUVI observations in Figure \ref{plot03}(f)--(g).
As illustrated in Figure \ref{fig:plot03sta}(d)--(f),
the bulged upper flank enters the FOV of COR1 and then propagates radially as the CME leading edge.
The STEREO A SECCHI observations bridge the FOV gap between GOES SUVI and SOHO LASCO C2,
which confirm that the upper flank in the non-radial stage bulges
and subsequently becomes the leading edge in the radial stage.
{In this context, the ``upper flank'' is defined as the outermost edge
that is farthest from the photosphere;
it is the flank on the greater-height side when the CME is in the non-radial stage;
it is distinct from the CME's northern edge in the radial stage.}

These observations reveal that the CME has undergone a non-radial to radial transition
before {leaving the FOV of STEREO A COR1 and entering the LASCO C2 FOV}.
The CME redirects itself by bulging its upper flank
{(the outermost edge in the non-radial stage)} toward the higher corona,
during which the CME's original leading edge in the non-radial stage
transforms into the southern flank in the radial stage.
As denoted by the cyan arrow in Figure \ref{plot03}(f)--(h),
{the erupted filament observed by GOES SUVI 284 \AA{}
is offset from the CME center during the bulging of the upper flank.
The observation times of SUVI 284 \AA{} images in Figure \ref{plot03}(g)--(h)
are nearly identical to those of AIA 304 \AA{} images in Figure \ref{plot03}(a)--(b),
and the observed filament material is spatially consistent at both wavelengths.
These observations indicate that, during the CME directional transition,
the filament is moving non-radially and is closer to the lower CME boundary near the photosphere.
The filament is also imaged by STEREO A
as shown in the EUVI 304 \AA{} and COR1 images with nearly simultaneous observation times
(Figure \ref{fig:plot03sta}(g)--(h)).
The CME core (``Core 1'') in COR1, south of the 0\textdegree{} latitude,
is roughly ahead of the filament in EUVI 304 \AA{},
which likely corresponds to the leading part of the filament,
while the trailing part is still in the view of EUVI.
The evolution of the filament in the FOV of COR1 is further analyzed in Appendix \ref{app:core}.
The further analysis reveals that a major portion
(probably the trailing part) of the filament
is displaced to the southern part of the CME after the directional transition,
which corresponds to the core (``Core 2'') in the images of STEREO A COR2
(Figure \ref{fig:plot03sta}(i)) and SOHO LASCO C2 (Figure \ref{plot03}(i)).
The connection between the filament and the CME core is established by only imaging observations,
which may allow for alternative explanations for ``Core 2''.
However, the displacement of the filament is a more consistent explanation for the CME core,
while a white-light bright core usually corresponds to the filament/prominence of a CME
\citep[see also][]{VourlidasLH2013SoPh,SongZC2019ApJ}.
During the transition from non-radial to radial,
the filament largely maintains its initial non-radial direction,
which deviates from the CME's overall propagation direction.
Consequently, a major portion of the filament is displaced,
and is observed as a white-light core (``Core 2'') in the southern part of the CME.}

{An animation for Figure \ref{plot03} is available,
where in the first part,
the SDO AIA 304 \AA{} movie illustrates that the filament moves almost horizontally
{(roughly parallel to the solar surface)}
and passes the 0\textdegree{} latitude around 18:15 UT on March 3.
The second part is from the running-difference images of GOES SUVI 284 \AA{} and SOHO LASCO C2,
which shows that the bulged upper CME flank observed by SUVI
is captured by LASCO C2 (marked by an arrow) around 18:11 UT
and then propagates nearly radially as the CME leading edge.}
{The northward expansion is also shown in this animation,
which is no more prominent than the southward expansion in the FOV of LASCO C2.
The animation for Figure \ref{fig:plot03sta} consists of
running-difference images of STEREO A EUVI 195 \AA{} and COR1, and images of COR2.
It illustrates that the upper flank in the non-radial stage bulges in the FOV of EUVI 195 \AA{}
and then enters the FOV of COR1 around 18:01 UT on March 3,
eventually propagating approximately radially as the CME leading edge.}

\subsection{CME on 2023 March 4}\label{sec:cme04}
\begin{figure*}[ht!]
  \centering
  \includegraphics[width=0.9\textwidth]{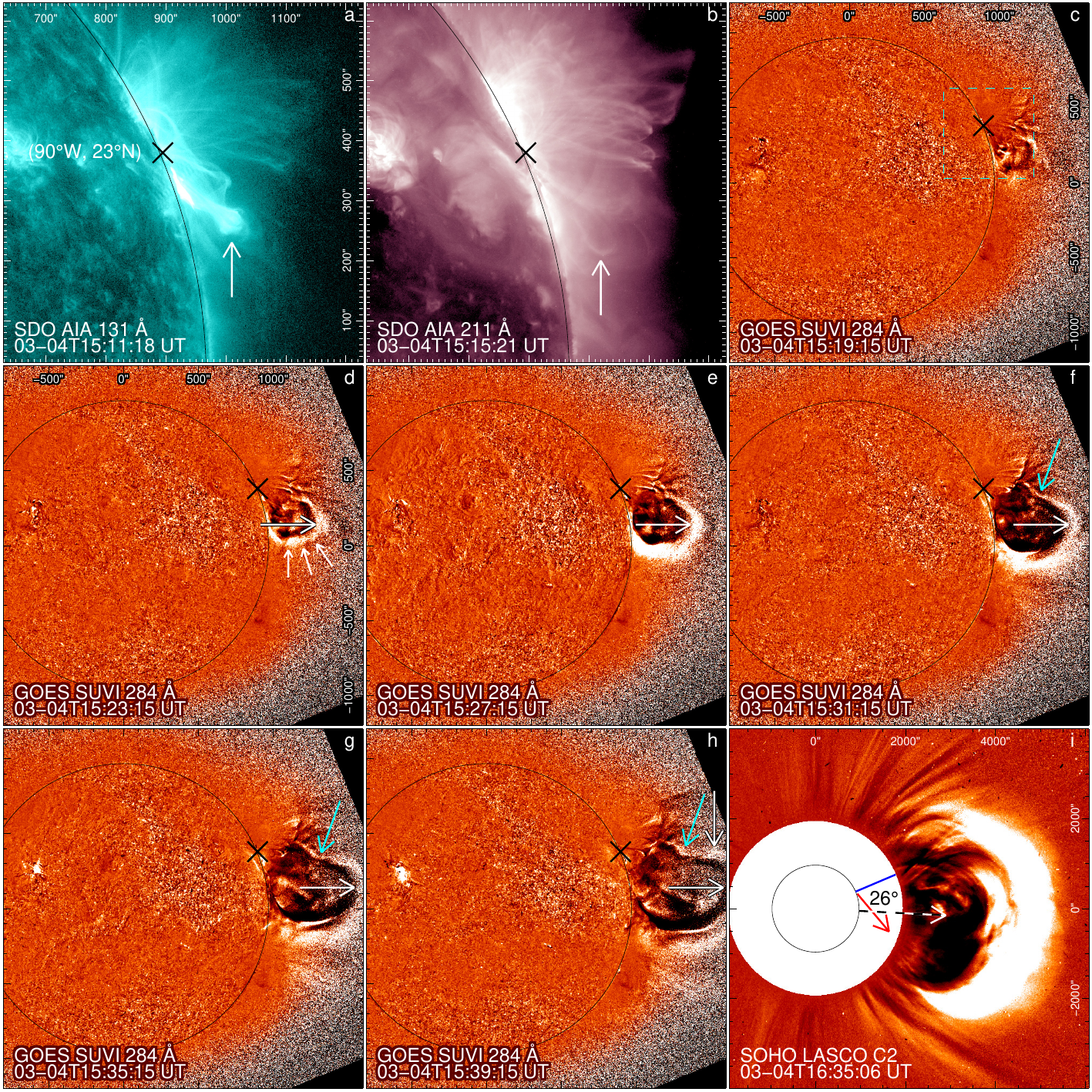}
  \caption{\label{plot04}The CME on 2023 March 4 {observed near the Earth}.
  (a)--(b) The erupted hot channel/flux rope observed by SDO AIA 131 \AA{}/211 \AA{},
  which is indicated by the white arrows.
  (c)--(h) The CME imaged by GOES SUVI 284 \AA{},
  where the arrows in (d) mark the CME leading edge {in the non-radial stage,
  the rightward white arrows in (d)--(h) denote the bulging upper CME flank},
  the cyan arrows in (f)--(h) indicate the indentation on the CME flank,
  {and the downward white arrow in (h) points to the northward expanding part.}
  (i) The CME observed in SOHO LASCO C2,
  where the red arrow, dashed arrow and blue line have the same meanings as in Figure \ref{plot03}(i).
  The symbol ``\textbf{\texttimes}'' in (a)--(h) marks the eruption site.
  The dashed square in (c) represents the field of view of (a)--(b).
  The coordinate axes are similar to those in Figure \ref{plot03}.
  {An animation for this figure is available, which consists of two parts.
  The first part corresponds to (b),
  which begins at 2023-03-04T15:05 UT and ends at 2023-03-04T15:33 UT.
  The second part is for (c)--(i),
  where the GOES SUVI images are superimposed on the LASCO C2 images.
  The second part runs from 2023-03-04T15:11 UT to 2023-03-04T16:35 UT,
  with an arrow indicating the CME leading edge in the LASCO C2 image at 15:35 UT.
  The real-time animation duration is $\sim$11 s.
  The annotations in the animation are similar to those in the animation for Figure \ref{plot03}. }}
  \end{figure*}

A similar transition of propagation direction also occurs for the CME on 2023 March 4 from the same active region.
This CME is associated with an M5.2 class flare that peaked at 15:57 UT on March 4 at (90\textdegree W,23\textdegree N).
Figure \ref{plot04}(a) shows a hot channel observed by SDO AIA 131 \AA{},
which is a typical signature of an erupting flux rope
\citep[e.g.,][]{ChengZD2013ApJ,HuLW2016ApJ}.
{The hot channel in the image appears elongated generally along its non-radial motion direction,
which indicates that the erupted flux rope is in an edge-on view,
otherwise the hot channel would be wider perpendicular to its motion direction.}
The flux rope is also observed in the AIA 211 \AA{} image as marked by the arrow in Figure \ref{plot04}(b),
which also moves in a non-radial direction similar to the erupted filament on March 3.
The feature indicated by the arrow is the cross-section of the flux rope {with a narrow cavity},
which is perpendicular to the line of sight.
The edge-on view suggests that the flux rope's apex is roughly aligned with the line of sight.
The narrow, compact cross-section resembles that of the well-known CME flux rope on 2017 September 10,
which is also on the solar limb and in an edge-on view \citep[e.g.,][]{YanYX2018ApJ}.
This flux-rope orientation is also consistent with the magnetic polarity inversion lines (PILs)
probably associated with the eruption, as well as the tilt-angle parameter in the GCS fitting.
The details of the GCS fitting and determining the orientation of the flux rope and the PILs
are provided in Appendix \ref{app:gcs} and \ref{app:orient}.

The CME is tracked by GOES SUVI after it leaves the FOV of SDO AIA,
as illustrated in Figure \ref{plot04}(c)--(h).
The upper flank of the CME, which is on the greater-height side
as marked by the rightward white arrows in Figure \ref{plot04}(d)--(h),
also bulges toward the higher corona.
Part of the flank is constrained from expanding,
which causes an indentation as indicated by the cyan arrows in Figure \ref{plot04}(f)--(h).
Similar to the CME on March 3,
the bulging of the upper flank of the CME on March 4
eventually leads to a transition from non-radial to radial propagation,
after which the CME's bulged upper flank becomes the new leading edge in the radial stage.
{It should be noted again that the ``upper flank'' here refers to the outermost edge of the CME
in the non-radial stage (as indicated by the rightward white arrows in Figure \ref{plot04}(d)--(h)),
and is not the northern edge in the radial stage.}
{The CME flank also expands northward at the location
marked by the downward white arrow in Figure \ref{plot04}(h).
However, as shown in the GCS fitting results in Appendix \ref{app:gcs},
this expansion does not cause a global northward deviation of the CME in the FOV of LASCO C2.}
The GCS fitting gives a final propagation direction of $-$3\textdegree{} latitude,
separated by $\sim$26\textdegree{} in latitude from the eruption site,
as shown in Figure \ref{plot04}(i).

\begin{figure*}[ht!]
  \centering
  \includegraphics[width=0.9\textwidth]{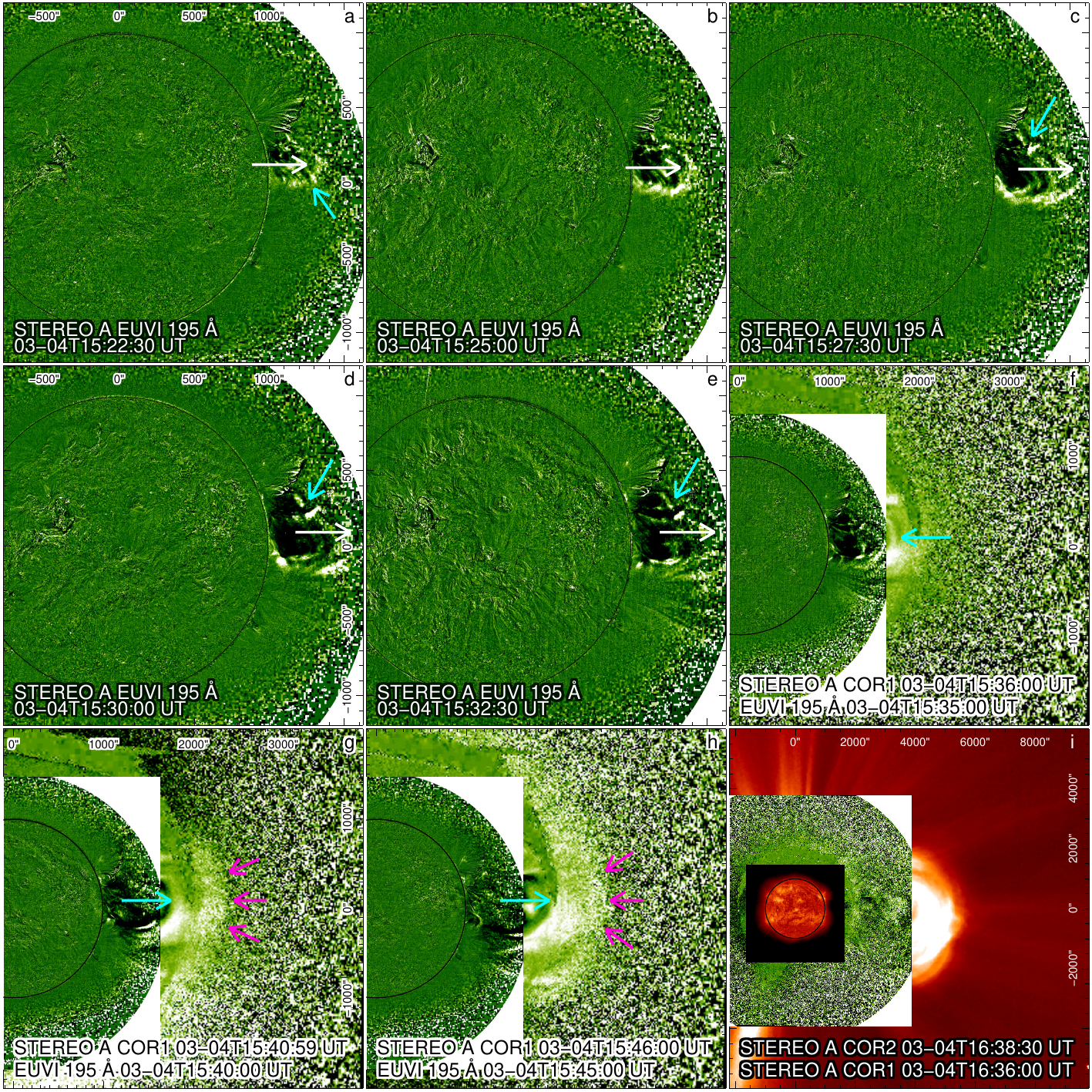}
  \caption{\label{fig:plot04sta}{The CME on 2023 March 4 observed by STEREO A.
  (a)--(e) Running-difference images of STEREO A EUVI 195 \AA{},
  where the rightward white arrows mark the bulging upper flank,
  the cyan arrow in (a) indicates the CME leading edge during the non-radial propagation,
  {and the cyan arrows in (c)--(e) point to the indentation on the CME flank,
  similar to those in Figure \ref{plot04}(f)--(h)}.
  (f)--(h) Running-difference images of STEREO A COR1 overlaid with running-difference images of EUVI 195 \AA{},
  where the cyan arrows denote the bulged flank appearing as the leading edge in the radial stage in COR1.
  The three pink arrows in (g)--(h) mark a plausible shock ahead of the radial leading edge,
  but the shock is not definitive due to the noisy imaging.
  (i) STEREO A COR2 image superimposed with a running-difference COR1 image and an EUVI 304 \AA{} image,
  showing the CME propagating nearly radially in COR2.
  The axes are labeled as in Figure \ref{plot03}.
  An animation for this figure is available,
  which has the same field of view as (i),
  where the COR2 images are superimposed with running-difference images of COR1 and EUVI 195 \AA{}.
  The animation runs from 2023-03-04T15:05 UT to 2023-03-04T17:17 UT,
  and the real-time duration is $\sim$3 s.
  The times are shown in the animation.}
  }\end{figure*}

{The CME on 2023 March 4 is also observed by STEREO A,
which is slightly behind the west limb of the Sun from STEREO A's perspective.
Figure \ref{fig:plot04sta}(a)--(e) display the running-difference images of STEREO A EUVI 195 \AA{},
where the cyan arrow in (a) indicates the CME's leading portion when the CME is propagating non-radially,
and the rightward white arrows in (a)--(e) mark the upper CME flank that is bulging toward the higher corona.
The cyan arrows in (c)--(e) point to an indentation on the CME flank,
which is similar to that observed by GOES SUVI in Figure \ref{plot04}(f)--(h).
Figure \ref{fig:plot04sta}(f)--(h) present the running-difference images of STEREO A COR1 overlaid with EUVI 195 \AA{},
where the horizontal cyan arrow in (f) denotes the bulged upper flank entering COR1.
The cyan arrows in (g)--(h) indicate the bulged flank propagating as the CME leading edge in COR1,
demonstrating that the CME has transitioned to approximately radial propagation in COR1 FOV.
The three pink arrows in (g)--(h) mark a plausible shock ahead of the CME leading edge in the radial stage.
However, the shock appears diffuse and lacks a sharp front, and is therefore not discussed further in this study.
Figure \ref{fig:plot04sta}(i) shows the CME propagating nearly radially in STEREO A COR2.
The STEREO A observations of both CMEs are consistent with those from near the Earth.}

{An animation for Figure \ref{plot04} is available,
where in the first part, the SDO AIA 211 \AA{} movie
shows the non-radial motion of the flux rope in an edge-on view.
The second part illustrates that
the bulged upper flank observed by GOES SUVI 284 \AA{}
appears in the view of LASCO C2 (indicated by an arrow) around 15:35 UT on March 4,
which then propagates almost radially as the CME leading edge, similar to the March 3 CME.}
{The animation for Figure \ref{fig:plot04sta} is similar to that for Figure \ref{fig:plot03sta}.
It also illustrates the transition of the CME from non-radial to radial propagation
by bulging its upper flank (on the greater-height side) toward the higher corona in the FOV of STEREO A EUVI 195 \AA{}.
After the transition, the CME propagates generally radially in the FOVs of STEREO A COR1 and COR2.}

\section{Ambient Magnetic Fields}\label{sec:fields}
\subsection{Overlying Magnetic Loops}\label{sec:loops}
\begin{figure*}[ht!]
  \centering
  \includegraphics[width=0.9\textwidth]{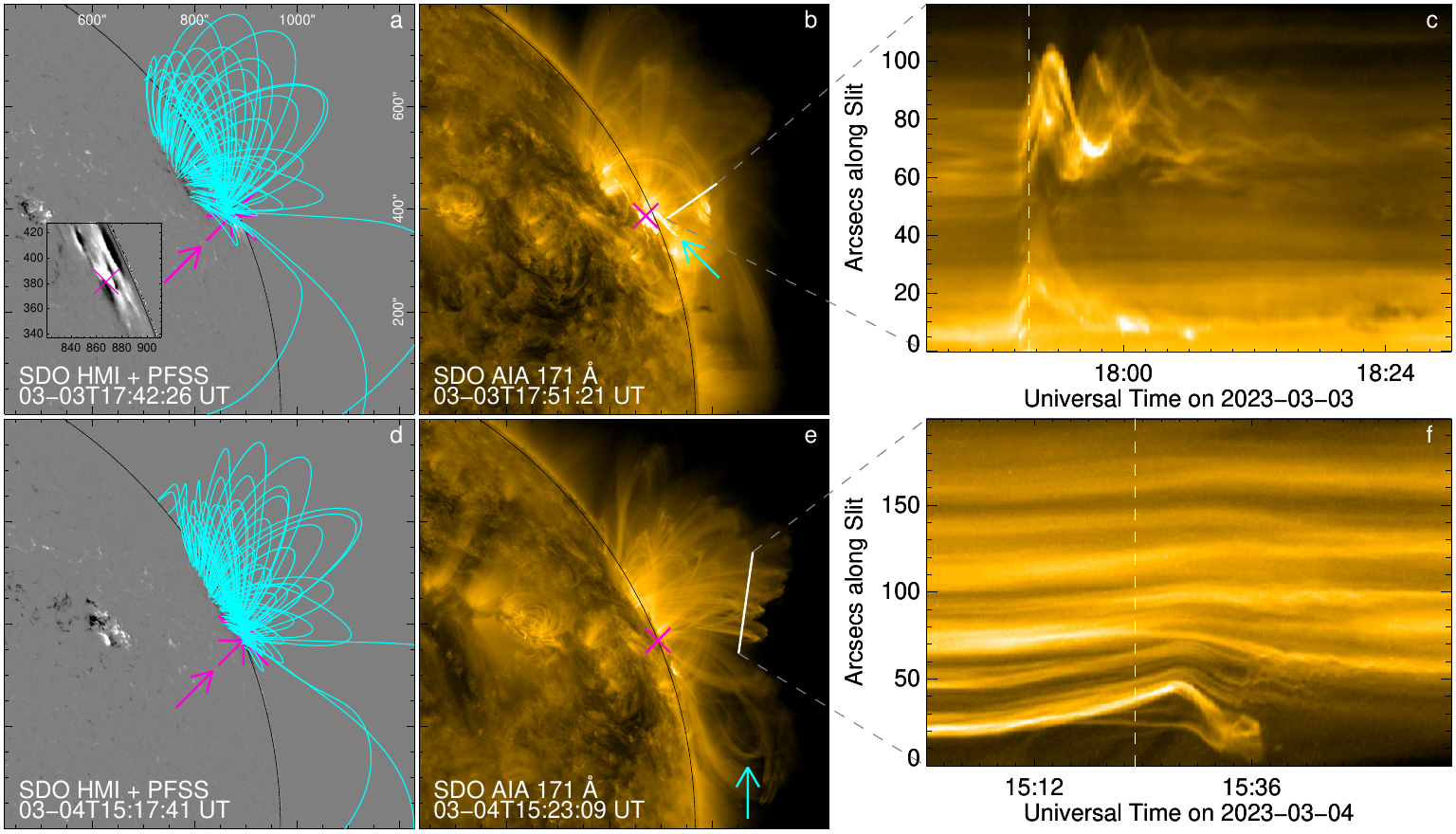}
  \caption{\label{pfss}The magnetic loops overlying the eruption site.
  (a) PFSS-extrapolated magnetic loops (cyan curves)
  superimposed on an SDO HMI line-of-sight magnetogram taken around the eruption on 2023 March 3,
  where the inset shows a zoomed-in view of the magnetogram
  centered on the eruption site (marked by the symbol ``\textbf{\texttimes}'').
  (b) SDO AIA 171 \AA{} image showing the overlying loops,
  where the symbol ``\textbf{\texttimes}'' denotes the eruption site,
  and the cyan arrow indicates the brightened erupting filament.
  (c) A time-distance profile along the slit marked by the white line in (b),
  where the vertical dashed line marks the image time in (b).
  (d) Similar to (a), showing the magnetic loops in the view around the eruption on 2023 March 4.
  (e) Similar to (b), but for the eruption on March 4,
  where the cyan arrow indicates the erupted flux rope in an edge-on view.
  (f) Similar to (c), but the time-distance profile is made along the slit in (e)
  where the dashed line marks the time in (e).
  The pink arrows in (a) and (d) point to the eruption site
  denoted by the symbol ``{\rlap{+}{\texttimes}}''.}
  \end{figure*}

The ambient magnetic field is reconstructed
using the Potential Field Source Surface (PFSS) model
based on a synoptic magnetogram at 18:03 UT on 2023 March 3,
which reveals a system of loops overlying both eruptions.
The loops are mapped on the magnetogram (Figure \ref{pfss}(a) and (d))
from the Helioseismic and Magnetic Imager \citep[HMI,][]{ScherrerSB2012SoPh} onboard SDO.
The PFSS-extrapolated magnetic loops are the same on March 3 and 4
but are viewed from slightly different perspectives owing to the rotation of the Sun.
The loops are also visible in the SDO AIA 171 \AA{} images in Figure \ref{pfss}(b) and (e),
as well as in images at other wavelengths displayed in Figure \ref{plot03} and \ref{plot04}.
As shown in the inset of Figure \ref{pfss}(a),
the eruption site of the March 3 CME is likely located at a PIL.
Figure \ref{pfss}(b) shows that the eruption site denoted by the symbol ``\textbf{\texttimes}''
is beneath the overlying loops and near the western footpoints of the loops,
which agrees with the HMI magnetogram taken on 2023 February 28
(see Figure \ref{fig:orient} in Appendix \ref{app:orient}).
{The narrow brightened filament in the AIA 171 \AA{} image on March 3
is elongated along the motion direction, as indicated by the arrow in Figure \ref{pfss}(b),
which} indicates that the erupted flux rope is in an edge-on view.
The erupting flux rope on 2023 March 4, marked by the arrow in Figure \ref{pfss}(e)
is similar to that in Figure \ref{plot04}(b),
which {exhibits a pattern with a narrow cavity
consistent with a flux-rope cross-section perpendicular to the line of sight}.
As displayed in the AIA 171 \AA{} image on March 4,
most of the observed loops with sharp edges reach a height of approximately 1.2 R$_\odot$,
and the most extended loops are visually estimated to 1.35 R$_\odot$.

For both events on 2023 March 3 and 4,
{each CME propagates non-radially beneath the overlying loops}.
After passing under the loops,
the CME bulges the upper flank and transitions to nearly radial propagation.
The expanding CME interacts with the loops during the directional transition.
The interaction causes transverse oscillations of the loops as if plucking musical instrument strings,
as illustrated in the time-distance profiles in Figure \ref{pfss}(c) and (f).
The oscillation on March 3 has been investigated by \citet{LiHB2024ScChE},
which also resembles another limb case with similar conditions to those in this study \citep{ZhangCL2022SoPh}.

The view on 2023 February 28 in Figure \ref{fig:orient} of Appendix \ref{app:orient}
shows that the overlying loops have a west-east orientation,
which is consistent with the AIA images in Figure \ref{pfss}(b) and (e).
This suggests that the loops are generally parallel to the flux-rope axis during the eruption,
and are not strapping over the flux rope.
The parallel overlying loops behave differently from strapping loops,
where the latter can affect the eruption of a flux rope
by governing the torus instability \citep[e.g.,][]{KliemT2006PhRvL,ChengZD2011ApJ}.
The magnetic fields around the eruption site are investigated further in the following paragraphs.

\subsection{Magnetic Field Properties}\label{sec:properties}
\begin{figure*}[ht!]
  \centering
  \includegraphics[width=0.9\textwidth]{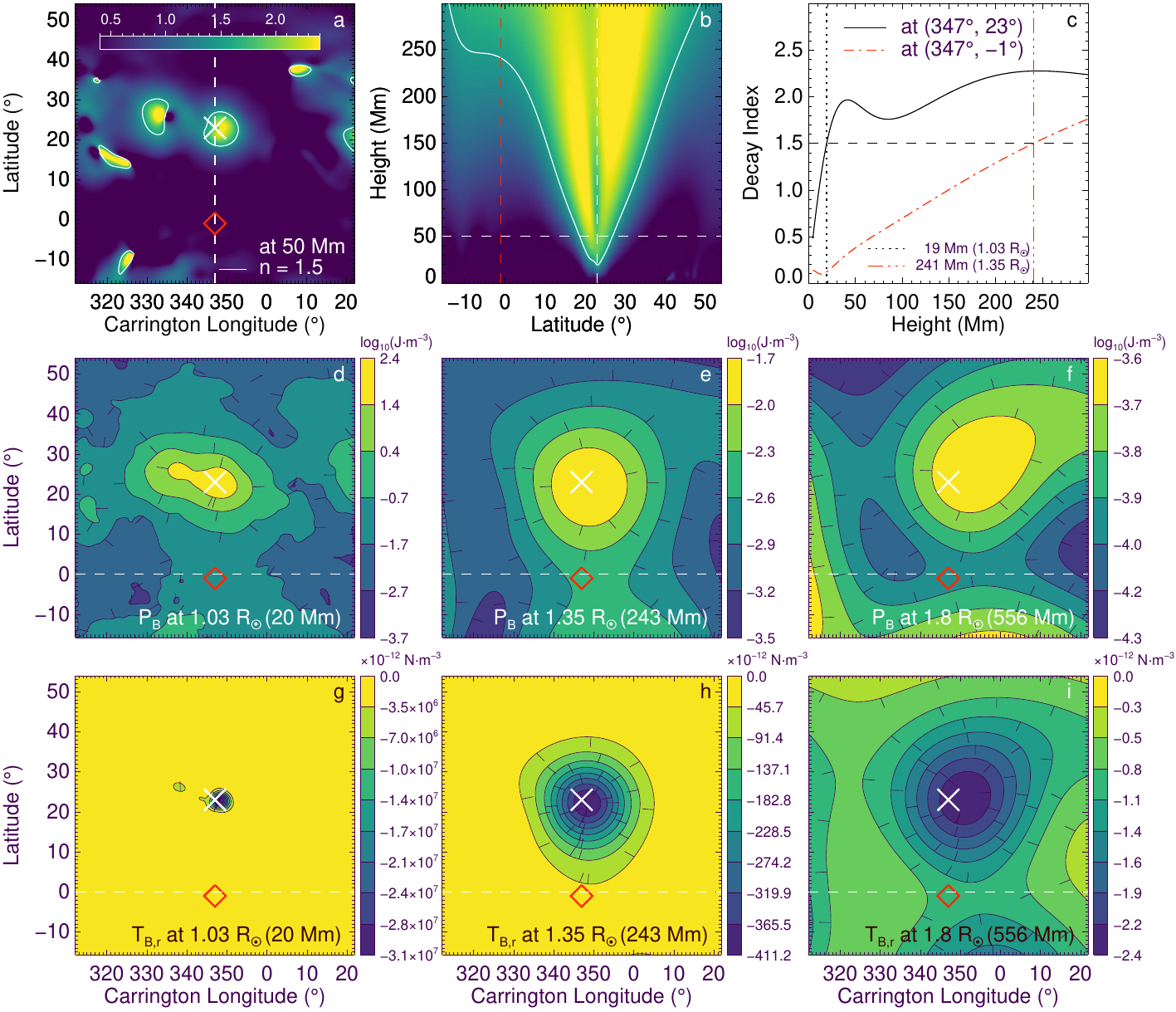}
  \caption{\label{dibe}Decay index $n$, magnetic energy density $P_{\mathrm{B}}$,
  and radial component of magnetic tension force $T_{\mathrm{B,r}}$ surrounding the eruption site.
  (a) The distribution of $n$ at the height of 50 Mm above the photosphere,
  where the white contour is $n = 1.5$,
  the vertical dashed line marks the longitude of 347\textdegree{} passing the eruption site,
  and the symbols ``\textbf{\texttimes}'' and ``\scalebox{1}[1]{$\Diamond$}'' mark the eruption site
  and the latitude corresponding to the CME radial propagation direction, respectively.
  (b) The distribution of $n$ in the meridional plane of 347\textdegree{} longitude,
  where the white contour is $n = 1.5$, the horizontal dashed line indicates the height for (a),
  and the orange-red and the white dashed vertical lines denote the latitudes
  for the symbols ``\scalebox{1}[1]{$\Diamond$}'' and ``\textbf{\texttimes}'' in (a), respectively.
  (c) The profiles of $n$ above the eruption site (solid curve)
  and along the radial direction of the CME propagation (orange-red dash-dotted curve),
  where the vertical dotted and dash-dot-dot-dotted lines indicate the respective heights at which $n = 1.5$.
  (d)--(f) The distribution of $P_{\mathrm{B}}$ at the heights of 20 Mm, 243 Mm, and 556 Mm above the photosphere, respectively.
  (g)--(i) The distribution of $T_{\mathrm{B,r}}$ at the same heights as in (d)--(f).
  The horizontal dashed line in (d)--(i) is an eye guide for 0\textdegree{} latitude,
  and the symbols ``\textbf{\texttimes}'' and ``\scalebox{1}[1]{$\Diamond$}'' have the same meaning as in (a).
  The plots in (a) and (b) share the same color bar.}
\end{figure*}

A decay index represents the rate of decrease in the strength of overlying magnetic fields with height above the photosphere,
where the magnetic strength is that of the component perpendicular to the flux-rope axis \citep{KliemT2006PhRvL}.
Although the overlying loops for the two CMEs in this study are generally parallel to the flux-rope axis during the eruption,
the decay index $n = - \frac{d \ln B_{\mathrm{T}}}{d \ln h}$ is still calculated based on the PFSS-extrapolated magnetic fields,
where $B_{\mathrm{T}}$ is the strength of transverse magnetic fields
and $h$ is the height above the photosphere.
Figure \ref{dibe}(a)--(b) show the distributions of the decay index at the height of 50 Mm above the photosphere
and in the meridional plane of the eruption site (347\textdegree{} Carrington longitude).
The distribution indicates that the decay index at 50 Mm
is larger than a typical critical value of 1.5 surrounding the eruption site,
while it is far below 1.5 in the region where the CME propagates radially ($\sim$$-1$ Carrington latitude).
Figure \ref{dibe}(c) illustrates that the critical height is only $\sim$19 Mm (1.03 R$_\odot$) above the eruption site,
but is $\sim$241 Mm (1.35 R$_\odot$) near the CME propagation direction.
If the torus instability is the major driver of the eruption,
this suggests that the flux rope is more likely to rise above the eruption site
than in the final CME propagation direction,
about 25\textdegree{} away (24\textdegree{} and 26\textdegree{} for the two CMEs respectively).
A decay index should be derived from the strapping fields perpendicular to the flux-rope axis,
and along the flux rope's likely trajectory.
In this study, the geometry and orientation of the flux rope cannot be precisely determined due to the limb view.
Thus, the profiles of the decay index obtained from the transverse fields,
which are roughly parallel to the flux-rope axis,
are only used for a qualitative discussion,
and are not a meaningful indicator of torus instability for the flux rope under these conditions.

The distribution of the magnetic pressure (energy density) $P_{\mathrm{B}}$
around the eruption site is also obtained and portrayed in Figure \ref{dibe}(d)--(f).
Figure \ref{dibe}(d)--(e) reveal that the magnetic pressure below 1.35 R$_\odot$
surrounding the eruption site is much larger than that in other regions,
where the stronger pressure overlaps with the overlying loops.
Figure \ref{dibe}(f) shows that there is a low-pressure valley at 1.8 R$_\odot$
which is aligned with the final radial propagation direction of the CME.
The valley of minimum $P_{\mathrm{B}}$ persists at all heights above 1.8 R$_\odot$ (not shown here).
This pattern of $P_{\mathrm{B}}$ distribution is reminiscent of previous studies
where CMEs are deflected toward regions of lower magnetic pressure
\citep[e.g.,][]{GuiSW2011SoPh,KayOE2015ApJ,HuLW2017ApJ}.
In our study, the two CMEs erupt in an almost horizontal direction
(parallel to the solar surface, as shown in Figure \ref{plot03}, \ref{plot04}, and \ref{pfss}),
which can be categorized as sideways CMEs \citep{McCauleySS2015SoPh}.
The non-radial eruption of the two CMEs may be caused by an asymmetric magnetic configuration
\citep[e.g.,][]{SunHL2012ApJ,PanasencoMV2013SoPh},
as the eruption site is located at the periphery of a strong field region
(see the magnetogram in Figure \ref{fig:orient}).
However, in the limb view, the details of the magnetic configuration
during the eruption cannot be resolved in this study.
As discussed previously, after leaving the overlying loops with a high magnetic pressure,
the two CMEs expand laterally toward the higher corona,
which results in a transition to radial propagation.
However, the high magnetic pressure is insufficient
to explain the constraint on radial expansion in the non-radial stage,
because the magnetic pressure force is upward as the magnetic field strength decreases with height in the corona.
Therefore, the magnetic tension force ($\mathbf{T}_{\mathrm{B}}$) from the overlying loops is also investigated.

The radial component of the magnetic tension force $T_{\mathrm{B,r}}$ is computed
based on the PFSS-extrapolated magnetic overlying fields.
The distribution of $T_{\mathrm{B,r}}$ is illustrated in Figure \ref{dibe}(g)--(i),
where the negative sign indicates a downward force.
Akin to the magnetic pressure $P_{\mathrm{B}}$,
$T_{\mathrm{B,r}}$ above the eruption site is also larger than that in surrounding regions.
The similarity between the two distributions of $P_{\mathrm{B}}$ and $T_{\mathrm{B,r}}$
suggests that the tension force is also associated with the overlying loops.
The magnetic tension force and pressure force are the two components of the Lorentz force,
$\mathbf{j} \times \mathbf{B} = \mathbf{T}_{\mathrm{B}} - \nabla P_{\mathrm{B}} $,
where $\mathbf{j}$ is the current density, and $\mathbf{B}$ is the magnetic field.
In the PFSS model, the current $\mathbf{j}$ in the corona is assumed to be zero,
which implies that the tension force and the pressure force calculated from the model are in balance.
For actual physical conditions in our case,
the downward radial tension force $T_{\mathrm{B,r}}$ must dominate the upward pressure force,
which yields a net downward force on the CME structure.
Although the axis of the flux rope at the apex is roughly parallel to the overlying loops,
the legs of the flux rope are mostly perpendicular to the loops in the non-radial stage
(refer to the schematic flux-rope axis in Figure \ref{fig:orient}(b)).
Therefore, the radial tension force $T_{\mathrm{B,r}}$
from the ``strapping'' loops can be exerted on the flux-rope legs beneath the loops,
{which constrains the portion of the CME with the legs from radial expansion
(i.e., lateral expansion relative to the non-radial propagation direction)}.
As illustrated in Figure \ref{dibe}(i), the radial tension force $T_{\mathrm{B,r}}$ at 1.8 R$_\odot$
also exhibits a valley around 0\textdegree{} latitude, like the magnetic pressure $P_{\mathrm{B}}$.
Once the CMEs move out of the system of overlying loops,
they expand upward in the region where the downward tension force $T_{\mathrm{B,r}}$ is weaker.
This lateral deformation eventually leads to the transition of propagation direction from non-radial to radial.

\section{Discussions and Conclusions}\label{sec:conclusion}
We have investigated two large-scale CMEs that erupted non-radially
from the active region NOAA AR 13234 near the solar limb on 2023 March 3 and 4, respectively,
and transitioned to a radial direction during their propagation in the low corona.
In the non-radial stage, the two CMEs move nearly horizontally {(parallel to the solar surface)},
as compact structures beneath a system of overlying loops.
After escaping the confinement by the strong tension force of the loops,
the CMEs laterally deform by bulging their upper flanks
({the outermost portion,} on the greater-height side in the non-radial stage) toward the higher corona,
which results in the transition of propagation direction.
This study provides the first detailed analysis
of the lateral deformation of CMEs during such a directional transition,
and contributes an essential component to the complete picture of CME evolution.
The findings of this work are summarized and discussed below.

\subsection{Directional Transition from the Lateral Deformation}\label{dis:transition}
Both CMEs transition from a non-radial to a radial direction
by laterally expanding their upper flanks,
which is investigated here for the first time by EUV imaging with a large FOV.
The two CMEs erupt non-radially near 23\textdegree{} latitude under a system of loops
and move in an almost horizontal direction relative to the solar surface,
and eventually redirect to nearly radial after leaving the loops.
The final propagation direction is close to the solar equatorial plane
which is about 25\textdegree{} away from the eruption site
(24\textdegree{} and 26\textdegree{} for the two CMEs, respectively).
The EUV images show that the upper flanks bulge toward the higher corona
when the CMEs leave the overlying loops,
which results in a lateral deformation of the CME structures.
{Meanwhile, a brief northward expansion is observed for both CMEs
but it does not cause a northward deviation in the CME propagation direction within the FOV of LASCO C2
as revealed by the GCS fitting (see Table \ref{tab:gcs} in Appendix \ref{app:gcs}).
The bulging of the upper CME flank} leads to the directional transition,
and the bulged flank of the CME in the non-radial stage becomes the leading edge in the radial stage.

CMEs are thought to have a self-similar evolution like an expanding ``solid body'' in the corona,
during which their geometry does not apparently change
\citep[e.g.,][]{Low1984ApJ,BalmacedaVS2020SoPh,Dai2022ApJ}.
However, in the early stage of propagation in the corona,
a CME evolves dramatically (e.g., expanding from a compact flux rope into a large-scale structure),
which can hardly maintain this self-similarity \citep[e.g.,][]{ZhangDH2004ApJ,CremadesIM2020AA}.
For the two CMEs in this study, the structure deforms during the directional transition,
and the leading edge in the radial stage is not the same portion as in the non-radial stage.
The geometry and kinematics of a CME modulate the properties of a CME-driven shock
\citep[e.g.][]{PatsourakosV2009ApJ,MaRG2011ApJ,LiberatoreLV2023ApJ,HuZL2024ApJ}.
Previous research proposes that non-radial CMEs are more likely to be linked to chromospheric Moreton-Ramsey waves
\citep{ZhengLL2023ApJ,ZhongCN2025ApJ}.
An unprecedented Ly$\alpha$ wave associated with the CME on 2023 March 3
is indeed detected in the chromosphere and transition region (Y. Zhou et al. 2026, in preparation).
These findings suggest that non-radial CMEs {influence} the properties
of the solar atmosphere differently from radial CMEs.
{For instance, a non-radially propagating CME may drive a shock/wave
whose non-radial portion of the front is more compressed,
or be more likely to impact the lower solar atmosphere than a radial CME.}

In the initial propagation stage,
the CME speed usually exhibits a slow-rise phase followed by a significant acceleration phase
synchronized with the flare intensity \citep[e.g.,][]{ZhangDH2004ApJ,ChengZK2020ApJ},
which manifests that magnetic reconnection dominates the fast acceleration \citep[e.g.,][]{ZhuQL2020ApJ,RevaLB2024SoPh}.
Magnetic reconnection is also coupled with multiple mechanisms related to CME acceleration
\citep[e.g., torus instability, hoop force, and pressure gradient; see][]
{ShenWF2012JGRA,Welsch2018SoPh,XingAC2024ApJ},
which implies the complexity of understanding CME kinematics.
In our study, the initial propagation direction of the two CMEs in the low corona
is {non-radial and almost parallel to the solar surface,
while} the final direction is nearly radial and about 25\textdegree{} away from the eruption (flare) site.
This indicates that the CMEs in the radial stage are far from their triggering reconnection,
and the roles of reconnection in the kinematics of both the non-radial and radial stages
are likely different but remain unclear.
Future studies on kinematics of such CMEs may clarify the kinematic differences
and the contribution of reconnection in CMEs with and without non-radial propagation.

{This study also adds a new instance where
the final direction of a CME with a non-radial propagation stage
can differ significantly from the source region.
The eventual propagation direction of both CMEs in this study
is up to 25\textdegree{} away from the eruption site
after transitioning from non-radial to radial propagation.
This finding underscores the challenge of accurately forecasting CME-induced space weather
based solely on low-corona observations.}

\subsection{{Partial} Constraint on the Radial Expansion}\label{dis:constraint}
The overlying loops are nearly parallel to and not strapping over the erupting flux rope,
but the strong magnetic tension force constrains the radial expansion of {part of} the CME
{by acting on the flux-rope legs} during the directional transition.
A system of loops are observed in EUV images above the eruption site,
which is consistent with the PFSS-extrapolated magnetic fields.
Both CMEs move beneath the loops in the non-radial stage,
and begin to expand radially
{(i.e., laterally relative to the non-radial direction)} as they leave the loops.
During the expansion, the CMEs interact with the loops and cause the loops to oscillate.

The upper flank of the CME on 2023 March 4 shows an indentation,
which suggests that the {radial} expansion of {part of} the CME is constrained
{during the directional transition}.
Although the overlying loops are approximately parallel to the apex axis of the flux rope,
the decay index is calculated from the transverse magnetic field.
The minimum height for the critical decay index of $n = 1.5$ is close to the eruption site,
which is about 25\textdegree{} away from the final CME propagation direction.
This suggests that the torus instability is not a major factor
determining the radial ascent of the CMEs during the directional transition in this case,
when the general orientation of the overlying fields is parallel to the flux rope.
The magnetic pressure above the eruption site is stronger and coincident with the overlying loops,
but it is unlikely to constrain the radial expansion because this pressure force is upward.

The radial component of the magnetic tension force above the eruption site,
which is downward, is larger than that in the CME propagation direction.
Considering the flux rope is almost horizontal beneath the overlying loops in the non-radial stage,
the legs of the flux rope are nearly perpendicular to and strapped by the loops
(see Figure \ref{fig:orient}(b) in Appendix \ref{app:orient}).
Thus, the strapping force from the loops can be exerted on the legs,
and constrains the radial expansion of the flux rope {in the early non-radial stage}.
Once the CME squeezes out of the loops and the downward tension force becomes weaker,
the CME laterally expands by bulging its upper flank
{and the part containing the legs under the loops remains constrained,
which results in the} transition to radial propagation.
The tension force here is from the PFSS-extrapolated magnetic fields,
which may not be the only factor determining the expansion of the flux rope.
The forces from the magnetic field and current of the flux rope
are also crucial to the radial expansion of a CME
\citep[e.g.,][]{KliemT2006PhRvL,MyersYJ2015Natur,ZhongD2021NatCo},
which, however, are difficult to examine due to the limb view in this study.
Nevertheless, our results illuminate the lateral deformation of CMEs
and its connection to the overlying loops
during the transition from non-radial to radial in the low corona.

\subsection{Displacement of the Filament}\label{dis:filament}
{A major part of the} erupted filament is displaced after the directional transition for one of the CMEs.
{Although this conclusion is drawn from only imaging observations,
which can hardly rule out all other explanations,
the displacement of the filament is a more plausible explanation based on our analysis.}
The March 3 CME filament moves almost horizontally
with respect to the solar surface after the eruption,
as seen in the EUV images (Figure \ref{plot03}--\ref{fig:plot03sta}).
A filament usually appears as the bright core of a CME in coronagraph images
as suggested in previous studies \citep[e.g.,][]{VourlidasLH2013SoPh,SongZC2019ApJ}.
After the CME transitions to radial propagation,
{a major part of the} filament is displaced to the southern part of the CME,
{which corresponds to the CME core (``Core 2'' in Figure \ref{plot03}(i) and \ref{fig:plot03sta}(i))}.
The CME laterally expands its upper flank during the directional transition,
and the original leading edge in the non-radial stage becomes the new southern edge in the radial stage.
The expansion slightly affects the dense and compact filament at the bottom of the flux rope,
and the filament changes its direction reluctantly
because its inertia maintains its initial non-radial motion.
{Also, the filament may not be fully ionized
and therefore cannot be entirely guided by the CME magnetic fields.
Consequently, a major portion of} the filament lags
behind the {main body} of the CME during the directional transition,
and is displaced to the southern part, moving away from the ecliptic plane,
placing it closer to the original leading edge of the non-radial stage
(see Figure \ref{plot03}(i) {and \ref{fig:plot03sta}(i)}).

Previous studies suggest that the acceleration of a filament is usually lower than that of the CME leading edge,
which causes a speed difference between them in the corona
\citep[e.g.,][]{MaricicVR2009SoPh,MajumdarDM2024ApJ}.
Our results indicate that, for a CME with non-radial propagation,
there is also a potential difference in direction between the filament and the CME's leading edge in the corona.
Moreover, this suggests that self-similarity is not maintained during the early evolution of such CMEs.
Additionally, the displacement {of the filament exemplifies how
a filament can be offset from the ecliptic plane,
thus preventing its detection near the ecliptic plane.
This} may partly explain why some interplanetary CMEs lack an in situ filament signature at 1 au
even though they originate from filament eruptions
\citep[e.g.,][]{LepriZ2010ApJ,WangFZ2018AA},
given that non-radial propagation of CMEs is not uncommon.

\begin{acknowledgments}
  We thank Dr. James Klimchuk for insightful discussions at the 246th AAS meeting,
  which significantly improved this work.
  This work is supported by
  the Strategic Priority Research Program of the Chinese Academy of Sciences (No. XDB0560000),
  the National Natural Science Foundation of China
  (Nos. 42274201, 42150105, and 42204176),
  the National Key R\&D Program of China (Nos. 2022YFF0503800 and 2021YFA0718600),
  and the Specialized Research Fund for State Key Laboratories of China.
  C.C. acknowledges support from the Research Foundation
  of Education Bureau of Hunan Province of China (No. 23B0593).
  X.Z. is also supported by the China Meteorological Administration ``Space Weather Monitoring and Alerting''
  Key Innovation Team (CMA2024ZD01).
  We thank Prof. Rui Liu at USTC again for sharing the code
  to improve the drawing of the PFSS-extrapolated magnetic field lines.
  SDO is the first mission launched for NASA's Living With a Star Program.
  SOHO is a project of international cooperation between ESA and NASA.
  SUTRI is a collaborative project conducted by the National Astronomical Observatories of CAS,
  Peking University, Tongji University,
  Xi'an Institute of Optics and Precision Mechanics of CAS,
  and the Innovation Academy for Microsatellites of CAS.
  We acknowledge the use of data from GOES SUVI.
\end{acknowledgments}

\facilities{SDO, SOHO, GOES, SUTRI}
\software{SolarSoftWare (\citealt{FreelandH2012ssw}), JHelioviewer (\citealt{MuellerNF2017AA})}

\appendix \label{appendix}
\section{Radial Propagation Direction} \label{app:gcs}
The propagation direction of the CME in the FOV of the SOHO LASCO C2
is determined by the GCS model \citep{ThernisienHV2006ApJ},
{combined with the images of STEREO A COR2}.
The GCS model approximates the morphology of a CME with a croissant-like structure,
which has six parameters: the longitude, latitude, tilt angle, height, aspect ratio, and half angle.
These parameters are described in detail in \citet{ThernisienHV2006ApJ}.
With multi-viewpoint observations,
the GCS model can fit the propagation direction and kinematics in addition to the morphology.
In this study,
STEREO A is located at a small separation angle (about 12\textdegree{}) from SOHO,
which does not provide a significant advantage in constructing a two-viewpoint observation with SOHO.
{However, the GCS fitting is based on observations from SOHO LASCO and STEREO A,
from which the CMEs are both viewed as limb events.}
For limb CMEs observed from a single viewpoint,
to obtain a reliable propagation direction in latitude,
the parameters longitude, tilt angle, aspect ratio, and half angle are reasonably fixed,
ignoring possible changes in longitude.

{Figure \ref{fig:gcs_c2_example} presents an example of GCS fitting for each CME
{based on images of STEREO A COR2 and LASCO C2},
and its animation illustrates all the fittings {in the view of LASCO C2}.
The GCS parameters for the two CMEs are given in Table \ref{tab:gcs},
where the latitude and height are fitted at each time.
The uncertainties of latitude and height in the table are around 0.1\textdegree{} and 0.1 R$_\odot$, respectively,
which are estimated from multiple fittings with other parameters fixed.
However, when fitting all free parameters,
the uncertainties can reach at least 6\textdegree{} and 0.6 R$_\odot$,
or even larger for single-viewpoint observations,
as suggested in a quantitative analysis by
\citet{VerbekeMK2023AdSpR}.
The fixed tilt angle of $-$20\textdegree{} in this study
is estimated from the general orientation of the PILs (see Appendix \ref{app:orient}).
It is consistent with the edge-on view of the erupted flux rope,
although the uncertainty can be up to 25\textdegree{} and subject to personal preferences
\citep{KayP2024SpWea}.
Nonetheless, the GCS fitting is adequate to determine
the final latitudinal propagation direction of the limb CMEs in this study,
despite the uncertainties.}

Our GCS fitting results reveal that,
in the later stages within the LASCO C2 FOV,
the propagation latitude of each CME stabilizes
near the equatorial plane ($-1$\textdegree{} and $-3$\textdegree{}, respectively).
Given this proximity, a representative latitude of $-1$\textdegree{} is adopted
to discuss the environmental magnetic field of both events,
as marked by the ``\scalebox{1}[1]{$\Diamond$}'' symbol in Figure \ref{dibe}.

{\begin{figure*}
  \centering
  \includegraphics[width=0.8\textwidth]{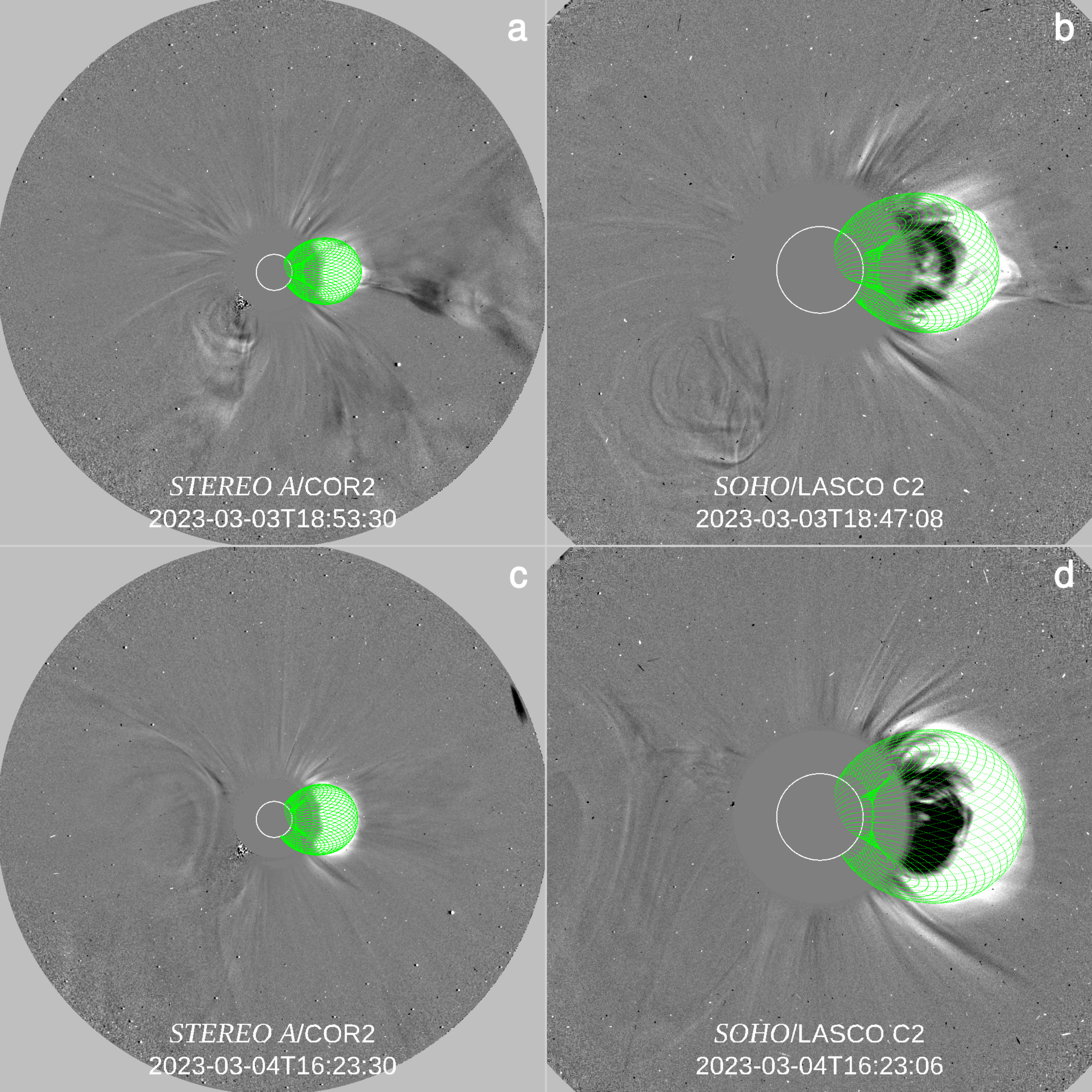}
  \caption{\label{fig:gcs_c2_example}GCS model fitting examples for the CMEs
  {on 2023 March 3 ((a)--(b)) and March 4 ((c)--(d))}, respectively.
  The overlaid green mesh represents the modeled CME geometry
  corresponding to {running-difference images of STEREO A COR2 and SOHO LASCO C2}.
  An animation with two panels for this figure is available,
  which shows all the GCS fittings {in the view of LASCO C2}
  in accordance with the parameters in Table \ref{tab:gcs}.
  The left panel of the animation corresponds to {(b)} and runs from 2023-03-03T18:23 UT to 2023-03-03T19:23 UT.
  The right panel is for {(d)}, which begins at 2023-03-04T15:59 UT and ends at 2023-03-04T16:47 UT.
  The real-time animation duration is $\sim$2 s.}
\end{figure*}}

\begin{table*}[ht!]
  \centering
  \caption{GCS Parameters \label{tab:gcs}}
    \begin{tabular}{l c c c c c c}
      \hline
      \hline
      \textbf{Time} & Longitude & \textbf{Latitude}                  & Tilt Angle    & \textbf{Height} & Ratio & Half Angle    \\
      (UT)          & \multicolumn{2}{c}{(Carrington, \textdegree)}  & (\textdegree) & (R$_\odot$)     &       & (\textdegree) \\
      \hline
      \multicolumn{7}{l}{CME on 2023 March 3} \\
      18:23 & 347 &  9   & $-20$ & 3.0 & 0.5 & 30 \\
      18:35 & 347 &  5   & $-20$ & 3.5 & 0.5 & 30 \\
      18:47 & 347 &  2   & $-20$ & 4.2 & 0.5 & 30 \\
      18:59 & 347 &  0   & $-20$ & 4.9 & 0.5 & 30 \\
      19:11 & 347 & $-1$ & $-20$ & 5.6 & 0.5 & 30 \\
      19:23 & 347 & $-1$ & $-20$ & 6.2 & 0.5 & 30 \\
      \hline
      \multicolumn{7}{l}{CME on 2023 March 4} \\
      15:59 & 347 &  2   & $-20$ & 3.2 & 0.6 & 35 \\
      16:11 & 347 &  2   & $-20$ & 3.9 & 0.6 & 35 \\
      16:23 & 347 &  0   & $-20$ & 4.7 & 0.6 & 35 \\
      16:35 & 347 & $-3$ & $-20$ & 5.4 & 0.6 & 35 \\
      16:47 & 347 & $-3$ & $-20$ & 6.1 & 0.6 & 35 \\
      \hline
      \hline
    \end{tabular}
  \tablecomments{{The times correspond to the LASCO C2 images used for the fitting.}
  Only latitude and height are fitted for each time,
  whose uncertainties can be about 1\textdegree{} and 0.1 R$_\odot$, respectively.}
\end{table*}

\section{Flux-rope Orientation} \label{app:orient}
\begin{figure}[ht!]
  \centering
  \includegraphics[width=0.7\textwidth]{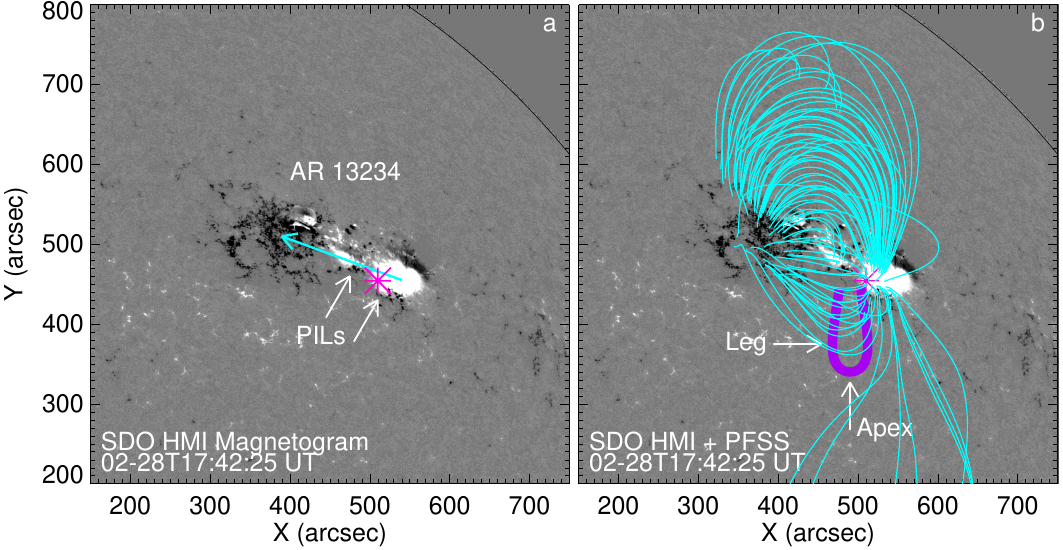}
  \caption{\label{fig:orient} Orientation of the flux rope and overlying loops.
  (a) An SDO HMI magnetogram from 2023 February 28,
  where white and black pixels are positive and negative polarities, respectively.
  The cyan arrow points from the positive to the negative polarity,
  indicating the general orientation of the overlying loops.
  The white arrows mark two polarity inversion lines (PILs) on the periphery of the positive polarity regions,
  which are potentially associated with the eruptions on 2023 March 3 and 4.
  (b) The same magnetogram overlaid with PFSS-extrapolated magnetic loops (cyan curves).
  The thick purple curve schematizes the flux-rope axis for both CMEs in the non-radial stage.
  The white arrows indicate the leg and apex of the flux rope.
  The symbol ``{\rlap{+}{\texttimes}}'' in both panels
  marks the approximate location of the eruptions with its position rotated back to February 28.}
\end{figure}

Due to the limb view in this study,
the orientation of the flux ropes can hardly be determined from the imaging observations.
Figure \ref{fig:orient}(a) shows an HMI magnetogram from 2023 February 28,
where the two white arrows indicate two PILs near the periphery of two positive polarity regions,
and the cyan arrow represents the generally horizontal orientation of the overlying loops.
Identifying the PIL for each eruption is difficult,
due to the strong projection effect in the limb view
and the expected evolution of the active region during the three days.
The eruptions on 2023 March 3 and 4 may be associated with one or both of the PILs.
Although the PILs are not accurately determined,
their major sections are roughly parallel to the overlying loops.
The ``{\rlap{+}{\texttimes}}'' symbol denotes an approximate location on February 28
corresponding to the eruption sites on March 3 and 4.
{The filament on March 3 is elongated roughly along its motion direction}
(Figure \ref{plot03}(a)--(c) and \ref{pfss}(b)),
which suggests that the erupted flux rope is in an edge-on view.
On March 4, the hot channel in the AIA 131 \AA{} image (Figure \ref{plot04}(a))
{is also elongated generally along its motion direction,
and the flux-rope cross-section in the AIA 211 \AA{} and 171 \AA{} images
exhibits a narrow cavity (Figure \ref{plot04}(b) and \ref{pfss}(e)).
These} indicate that the erupted flux rope is also in an edge-on view.
Therefore, the apex axis of each erupted flux rope is approximately parallel to the overlying loops,
which is also consistent with the potentially associated PILs in Figure \ref{fig:orient}(a).

In Figure \ref{fig:orient}(b),
the purple thick curve schematically represents the axis of the CMEs in the non-radial stage,
and the white arrows indicate the flux-rope leg and apex.
While the flux ropes are beneath the overlying loops,
their apex axes are nearly parallel to the loops and their legs are roughly perpendicular to the loops.
The horizontal white arrow can also represent the line of sight from the edge-on viewpoint for the CMEs.
Although the precise eruption sites are difficult to determine,
they must be close to each other and beneath the overlying loops
as marked by the ``{\rlap{+}{\texttimes}}'' symbol in Figure \ref{fig:orient}(b).
Note that the schematic purple curve
does not reflect the actual geometry and position of the erupted flux ropes.

{\section{Erupted Filament on March 3 Observed by STEREO COR1} \label{app:core}}

\begin{figure}[ht!]
  \centering
  \includegraphics[width=0.9\textwidth]{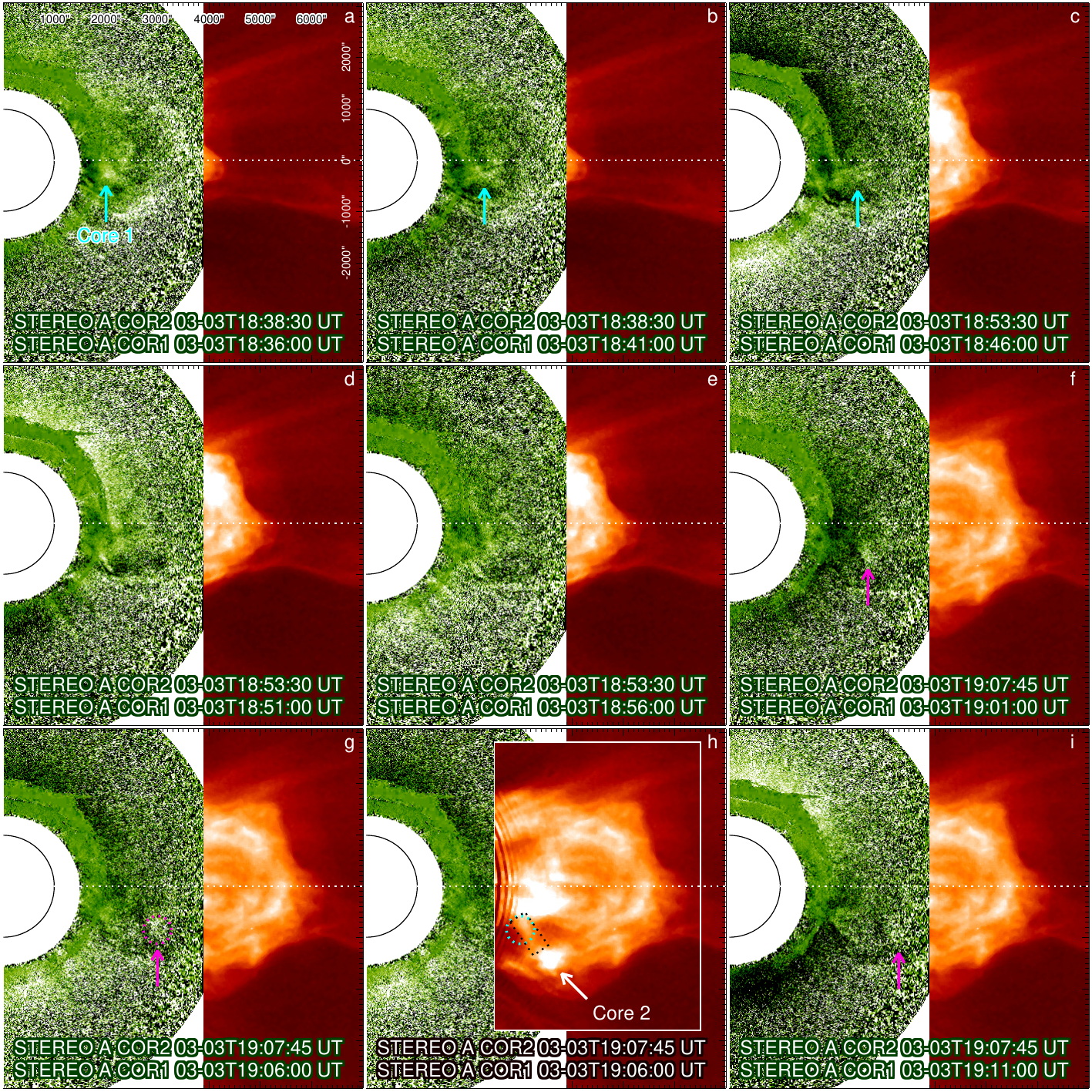}
  \caption{\label{fig:cor1cor2}{Erupted filament on 2023 March 3
  in the running-difference STEREO A COR1 images overlaid on STEREO A COR2 images.
  The cyan arrow labeled with ``Core 1'' in (a) indicates a CME core in COR1,
  which is the same as in Figure \ref{fig:plot03sta}(h)
  and corresponds to the leading part of the erupted filament.
  The cyan arrows in (b)--(c) mark the evolved CME core,
  which appears to be moving roughly radially on the south side of the ecliptic plane,
  but it is not discernible in the COR1 images in (d)--(e).
  The pink arrows in (f)--(g) and (i) denote a vague brightened feature moving along a more southward path.
  The image in (h) is identical to that in (g) but with part of the COR2 image overlaid,
  which shows another CME core (``Core 2'' indicated by the white arrow, also seen in Figure \ref{fig:plot03sta}(i)).
  The dotted circle in (g)--(h) represents the approximate position of the brightened feature in COR1.
  The dotted rectangle in (h) denotes a plausible structure connecting ``Core 2''.
  The axes are labeled in (a) similarly to those in Figure \ref{plot03}.
  The dotted line at a Y-position of 0 arcseconds approximately indicates the ecliptic plane.
  The brightened feature in COR1 (in (f)--(g) and (i)) and its motion are more visible in the animation for Figure \ref{fig:plot03sta}.}}
\end{figure}

{Figure \ref{fig:cor1cor2} presents running-difference images of STEREO A COR1 overlaid on STEREO A COR2,
which illustrates the connection between the erupted filament
and the white-light core in COR2 for the CME on 2023 March 3.
The cyan arrow in Figure \ref{fig:cor1cor2}(a) indicates a CME core (``Core 1'') in COR1,
which is the same as the one in Figure \ref{fig:plot03sta}(h).
This CME core is associated with the leading part of the elongated filament observed in STEREO A EUVI 304 \AA{}
as shown in Figure \ref{fig:plot03sta}(g)--(h) and as discussed in Section \ref{sec:cme03}.
Figure \ref{fig:cor1cor2}(b)--(c) suggest that ``Core 1'' plausibly moves approximately radially
on the south side of the ecliptic plane,
which is not identifiable in the subsequent COR1 images in Figure \ref{fig:cor1cor2}(d)--(e).
A vague brightened feature appears later in COR1
(indicated by the pink arrows in Figure \ref{fig:cor1cor2}(f)--(g) and (i)),
which also moves roughly radially but in a more southward direction.
The images in Figure \ref{fig:cor1cor2}(g)--(h) are identical,
where the CME core (``Core 2'') in COR2 is also shown in Figure \ref{fig:plot03sta}(i).
The dotted circle marks the approximate position of the brightened feature,
which spatially corresponds to a structure following and connecting ``Core 2''
(see the bright pixels inside the dotted rectangle in Figure \ref{fig:cor1cor2}(h)).
Although the brightened feature (the dotted circle) is not exactly coincident with ``Core 2'',
the structure connecting them in Figure \ref{fig:cor1cor2}(h) suggests
that they are likely linked to each other and form the same entity.
This entity is probably associated with the trailing part of the elongated filament observed at EUV wavelengths
(as shown in Figure \ref{plot03}(a)--(c) and Figure \ref{fig:plot03sta}(g)--(h)).
These observations further support the displacement of a major part of the filament after the directional transition
as discussed in Section \ref{sec:cme03} and \ref{dis:filament}.
The displaced filament is also consistent with the dense and bright core (``Core 2'') in the southern part of the CME.}

{The brightened feature in the running-difference COR1 images
is more visible in the animation for Figure \ref{fig:plot03sta}.
The filament-related entity is not clearly discernible in COR1 possibly due to the diffuse imaging,
and the connection between the filament and the CME core in COR2
may not be definitively established based only on white-light coronagraph observations.
However, based on the analysis above, a more probable scenario
for the CME core in COR2 (and also in LASCO C2)
is that it is associated with the displaced part of the erupted filament.
Moreover, the filament during its non-radial motion is already
offset from the CME center and closer to the southern boundary near the photosphere,
as observed in the EUV images (see Figure \ref{plot03}(a)--(b), (g)--(h), and the associated animation).}

\clearpage
\begin{thebibliography}{}
\expandafter\ifx\csname natexlab\endcsname\relax\def\natexlab#1{#1}\fi
\providecommand{\url}[1]{\href{#1}{#1}}
\providecommand{\dodoi}[1]{doi:~\href{http://doi.org/#1}{\nolinkurl{#1}}}
\providecommand{\doeprint}[1]{\href{http://ascl.net/#1}{\nolinkurl{http://ascl.net/#1}}}
\providecommand{\doarXiv}[1]{\href{https://arxiv.org/abs/#1}{\nolinkurl{https://arxiv.org/abs/#1}}}

\bibitem[{{Alobaid} {et~al.}(2023){Alobaid}, {Abduallah}, {Wang}, {Wang}, {Fan}, {Li}, {Cavus}, \& {Yurchyshyn}}]{AlobaidAW2023ApJ}
{Alobaid}, K.~A., {Abduallah}, Y., {Wang}, J. T.~L., {et~al.} 2023, \apjl, 958, L34, \dodoi{10.3847/2041-8213/ad0c4a}

\bibitem[{{Bai} {et~al.}(2023){Bai}, {Tian}, {Deng}, {Wang}, {Yang}, {Zhang}, {Zhang}, {Qi}, {Wang}, {Gao}, {Yu}, {He}, {Shen}, {Shen}, {Guo}, {Hou}, {Ji}, {Bi}, {Duan}, {Yang}, {Lin}, {Hu}, {Song}, {Yang}, {Chen}, {Qiao}, {Ge}, {Li}, {Jin}, {He}, {Chen}, {Zhu}, {He}, {Shi}, {Liu}, {Li}, {Xu}, {Liu}, {Li}, {Feng}, {Wang}, {Fan}, {Liu}, {Guo}, {Sun}, {Wu}, {Li}, {Yang}, {Ye}, {Gu}, {Wu}, {Zhang}, {Yu}, {Ye}, {Sheng}, {Wang}, {Li}, {Huang}, \& {Zhang}}]{BaiTD2023RAA}
{Bai}, X., {Tian}, H., {Deng}, Y., {et~al.} 2023, Research in Astronomy and Astrophysics, 23, 065014, \dodoi{10.1088/1674-4527/accc74}

\bibitem[{{Balmaceda} {et~al.}(2020){Balmaceda}, {Vourlidas}, {Stenborg}, \& {St. Cyr}}]{BalmacedaVS2020SoPh}
{Balmaceda}, L.~A., {Vourlidas}, A., {Stenborg}, G., \& {St. Cyr}, O.~C. 2020, \solphys, 295, 107, \dodoi{10.1007/s11207-020-01672-6}

\bibitem[{{C{\'e}cere} {et~al.}(2020){C{\'e}cere}, {Sieyra}, {Cremades}, {Mierla}, {Sahade}, {Stenborg}, {Costa}, {West}, \& {D'Huys}}]{CecereSC2020AdSpR}
{C{\'e}cere}, M., {Sieyra}, M.~V., {Cremades}, H., {et~al.} 2020, Advances in Space Research, 65, 1654, \dodoi{10.1016/j.asr.2019.08.043}

\bibitem[{{Chen} {et~al.}(2019){Chen}, {Liu}, {Wang}, {Zhao}, {Hu}, \& {Zhu}}]{ChenLW2019ApJ}
{Chen}, C., {Liu}, Y.~D., {Wang}, R., {et~al.} 2019, \apj, 884, 90, \dodoi{10.3847/1538-4357/ab3f36}

\bibitem[{{Chen} {et~al.}(2024){Chen}, {Liu}, {Zhu}, {Hu}, \& {Wang}}]{ChenLZ2024ApJ}
{Chen}, C., {Liu}, Y.~D., {Zhu}, B., {Hu}, H., \& {Wang}, R. 2024, \apjl, 969, L4, \dodoi{10.3847/2041-8213/ad53ca}

\bibitem[{{Cheng} {et~al.}(2011){Cheng}, {Zhang}, {Ding}, {Guo}, \& {Su}}]{ChengZD2011ApJ}
{Cheng}, X., {Zhang}, J., {Ding}, M.~D., {Guo}, Y., \& {Su}, J.~T. 2011, \apj, 732, 87, \dodoi{10.1088/0004-637X/732/2/87}

\bibitem[{{Cheng} {et~al.}(2013){Cheng}, {Zhang}, {Ding}, {Liu}, \& {Poomvises}}]{ChengZD2013ApJ}
{Cheng}, X., {Zhang}, J., {Ding}, M.~D., {Liu}, Y., \& {Poomvises}, W. 2013, \apj, 763, 43, \dodoi{10.1088/0004-637X/763/1/43}

\bibitem[{{Cheng} {et~al.}(2020){Cheng}, {Zhang}, {Kliem}, {T{\"o}r{\"o}k}, {Xing}, {Zhou}, {Inhester}, \& {Ding}}]{ChengZK2020ApJ}
{Cheng}, X., {Zhang}, J., {Kliem}, B., {et~al.} 2020, \apj, 894, 85, \dodoi{10.3847/1538-4357/ab886a}

\bibitem[{{Cremades} {et~al.}(2020){Cremades}, {Iglesias}, \& {Merenda}}]{CremadesIM2020AA}
{Cremades}, H., {Iglesias}, F.~A., \& {Merenda}, L.~A. 2020, \aap, 635, A100, \dodoi{10.1051/0004-6361/201936664}

\bibitem[{{Dai}(2022)}]{Dai2022ApJ}
{Dai}, X. 2022, \apj, 925, 24, \dodoi{10.3847/1538-4357/ac3eda}

\bibitem[{{Darnel} {et~al.}(2022){Darnel}, {Seaton}, {Bethge}, {Rachmeler}, {Jarvis}, {Hill}, {Peck}, {Hughes}, {Shapiro}, {Riley}, {Vasudevan}, {Shing}, {Koener}, {Edwards}, {Mathur}, \& {Timothy}}]{DarnelSB2022SpWea}
{Darnel}, J.~M., {Seaton}, D.~B., {Bethge}, C., {et~al.} 2022, Space Weather, 20, e2022SW003044, \dodoi{10.1029/2022SW00304410.1002/essoar.10510311.1}

\bibitem[{{Domingo} {et~al.}(1995){Domingo}, {Fleck}, \& {Poland}}]{DomingoFP1995SoPh}
{Domingo}, V., {Fleck}, B., \& {Poland}, A.~I. 1995, \solphys, 162, 1, \dodoi{10.1007/BF00733425}

\bibitem[{{Freeland} \& {Handy}(2012)}]{FreelandH2012ssw}
{Freeland}, S.~L., \& {Handy}, B.~N. 2012, {SolarSoft: Programming and data analysis environment for solar physics}, Astrophysics Source Code Library, record ascl:1208.013.
\newblock \doeprint{1208.013}

\bibitem[{{Gandhi} {et~al.}(2024){Gandhi}, {Patel}, {Pant}, {Majumdar}, {Pal}, {Banerjee}, \& {Morgan}}]{GandhiPP2024SpWea}
{Gandhi}, H., {Patel}, R., {Pant}, V., {et~al.} 2024, Space Weather, 22, e2023SW003805, \dodoi{10.1029/2023SW003805}

\bibitem[{{Gopalswamy} {et~al.}(2000){Gopalswamy}, {Lara}, {Lepping}, {Kaiser}, {Berdichevsky}, \& {St. Cyr}}]{GopalswamyLL2000GeoRL}
{Gopalswamy}, N., {Lara}, A., {Lepping}, R.~P., {et~al.} 2000, \grl, 27, 145, \dodoi{10.1029/1999GL003639}

\bibitem[{{Gopalswamy} {et~al.}(2009){Gopalswamy}, {M{\"a}kel{\"a}}, {Xie}, {Akiyama}, \& {Yashiro}}]{GopalswamyMX2009JGRA}
{Gopalswamy}, N., {M{\"a}kel{\"a}}, P., {Xie}, H., {Akiyama}, S., \& {Yashiro}, S. 2009, Journal of Geophysical Research (Space Physics), 114, A00A22, \dodoi{10.1029/2008JA013686}

\bibitem[{{Gopalswamy} {et~al.}(2001){Gopalswamy}, {Yashiro}, {Kaiser}, {Howard}, \& {Bougeret}}]{GopalswamyYK2001ApJ}
{Gopalswamy}, N., {Yashiro}, S., {Kaiser}, M.~L., {Howard}, R.~A., \& {Bougeret}, J.~L. 2001, \apjl, 548, L91, \dodoi{10.1086/318939}

\bibitem[{{Gui} {et~al.}(2011){Gui}, {Shen}, {Wang}, {Ye}, {Liu}, {Wang}, \& {Zhao}}]{GuiSW2011SoPh}
{Gui}, B., {Shen}, C., {Wang}, Y., {et~al.} 2011, \solphys, 271, 111, \dodoi{10.1007/s11207-011-9791-9}

\bibitem[{{Howard} {et~al.}(2008){Howard}, {Moses}, {Vourlidas}, {Newmark}, {Socker}, {Plunkett}, {Korendyke}, {Cook}, {Hurley}, {Davila}, {Thompson}, {St Cyr}, {Mentzell}, {Mehalick}, {Lemen}, {Wuelser}, {Duncan}, {Tarbell}, {Wolfson}, {Moore}, {Harrison}, {Waltham}, {Lang}, {Davis}, {Eyles}, {Mapson-Menard}, {Simnett}, {Halain}, {Defise}, {Mazy}, {Rochus}, {Mercier}, {Ravet}, {Delmotte}, {Auchere}, {Delaboudiniere}, {Bothmer}, {Deutsch}, {Wang}, {Rich}, {Cooper}, {Stephens}, {Maahs}, {Baugh}, {McMullin}, \& {Carter}}]{HowardMV2008SSRv}
{Howard}, R.~A., {Moses}, J.~D., {Vourlidas}, A., {et~al.} 2008, \ssr, 136, 67, \dodoi{10.1007/s11214-008-9341-4}

\bibitem[{Howard(2014)}]{Howard2014}
Howard, T. 2014, Introduction (New York, NY: Springer New York), 1--14, \dodoi{10.1007/978-1-4614-7975-8_1}

\bibitem[{{Hu} {et~al.}(2016){Hu}, {Liu}, {Wang}, {M{\"o}stl}, \& {Yang}}]{HuLW2016ApJ}
{Hu}, H., {Liu}, Y.~D., {Wang}, R., {M{\"o}stl}, C., \& {Yang}, Z. 2016, \apj, 829, 97, \dodoi{10.3847/0004-637X/829/2/97}

\bibitem[{{Hu} {et~al.}(2017){Hu}, {Liu}, {Wang}, {Zhao}, {Zhu}, \& {Yang}}]{HuLW2017ApJ}
{Hu}, H., {Liu}, Y.~D., {Wang}, R., {et~al.} 2017, \apj, 840, 76, \dodoi{10.3847/1538-4357/aa6d54}

\bibitem[{{Hu} {et~al.}(2024){Hu}, {Zhu}, {Liu}, {Chen}, {Wang}, \& {Zhao}}]{HuZL2024ApJ}
{Hu}, H., {Zhu}, B., {Liu}, Y.~D., {et~al.} 2024, \apj, 976, 9, \dodoi{10.3847/1538-4357/ad7ead}

\bibitem[{{Kaiser} {et~al.}(2008){Kaiser}, {Kucera}, {Davila}, {St. Cyr}, {Guhathakurta}, \& {Christian}}]{KaiserKD2008SSRv}
{Kaiser}, M.~L., {Kucera}, T.~A., {Davila}, J.~M., {et~al.} 2008, \ssr, 136, 5, \dodoi{10.1007/s11214-007-9277-0}

\bibitem[{{Kay} {et~al.}(2015){Kay}, {Opher}, \& {Evans}}]{KayOE2015ApJ}
{Kay}, C., {Opher}, M., \& {Evans}, R.~M. 2015, \apj, 805, 168, \dodoi{10.1088/0004-637X/805/2/168}

\bibitem[{{Kay} \& {Palmerio}(2024)}]{KayP2024SpWea}
{Kay}, C., \& {Palmerio}, E. 2024, Space Weather, 22, e2023SW003796, \dodoi{10.1029/2023SW003796}

\bibitem[{{Kliem} \& {T{\"o}r{\"o}k}(2006)}]{KliemT2006PhRvL}
{Kliem}, B., \& {T{\"o}r{\"o}k}, T. 2006, \prl, 96, 255002, \dodoi{10.1103/PhysRevLett.96.255002}

\bibitem[{{Lemen} {et~al.}(2012){Lemen}, {Title}, {Akin}, {Boerner}, {Chou}, {Drake}, {Duncan}, {Edwards}, {Friedlaender}, {Heyman}, {Hurlburt}, {Katz}, {Kushner}, {Levay}, {Lindgren}, {Mathur}, {McFeaters}, {Mitchell}, {Rehse}, {Schrijver}, {Springer}, {Stern}, {Tarbell}, {Wuelser}, {Wolfson}, {Yanari}, {Bookbinder}, {Cheimets}, {Caldwell}, {Deluca}, {Gates}, {Golub}, {Park}, {Podgorski}, {Bush}, {Scherrer}, {Gummin}, {Smith}, {Auker}, {Jerram}, {Pool}, {Soufli}, {Windt}, {Beardsley}, {Clapp}, {Lang}, \& {Waltham}}]{LemenTA2012SoPh}
{Lemen}, J.~R., {Title}, A.~M., {Akin}, D.~J., {et~al.} 2012, \solphys, 275, 17, \dodoi{10.1007/s11207-011-9776-8}

\bibitem[{{Lepri} \& {Zurbuchen}(2010)}]{LepriZ2010ApJ}
{Lepri}, S.~T., \& {Zurbuchen}, T.~H. 2010, \apjl, 723, L22, \dodoi{10.1088/2041-8205/723/1/L22}

\bibitem[{{Li} {et~al.}(2024{\natexlab{a}}){Li}, {Hou}, {Bai}, {Li}, {Fang}, {Zhao}, {Wang}, \& {Ning}}]{LiHB2024ScChE}
{Li}, D., {Hou}, Z., {Bai}, X., {et~al.} 2024{\natexlab{a}}, Science in China E: Technological Sciences, 67, 1592, \dodoi{10.1007/s11431-023-2534-8}

\bibitem[{{Li} {et~al.}(2024{\natexlab{b}}){Li}, {Jing}, {Song}, {Li}, {Tian}, {Liu}, {Wang}, {Ding}, {Battaglia}, {Feng}, {Li}, \& {Gan}}]{LiJS2024ApJ}
{Li}, Y., {Jing}, Z., {Song}, D.-C., {et~al.} 2024{\natexlab{b}}, \apjl, 963, L3, \dodoi{10.3847/2041-8213/ad27ca}

\bibitem[{{Liberatore} {et~al.}(2023){Liberatore}, {Liewer}, {Vourlidas}, {Braga}, {Velli}, {Panasenco}, {Telloni}, \& {Mancuso}}]{LiberatoreLV2023ApJ}
{Liberatore}, A., {Liewer}, P.~C., {Vourlidas}, A., {et~al.} 2023, \apj, 957, 110, \dodoi{10.3847/1538-4357/acf8bf}

\bibitem[{{Liewer} {et~al.}(2015){Liewer}, {Panasenco}, {Vourlidas}, \& {Colaninno}}]{LiewerPV2015SoPh}
{Liewer}, P., {Panasenco}, O., {Vourlidas}, A., \& {Colaninno}, R. 2015, \solphys, 290, 3343, \dodoi{10.1007/s11207-015-0794-9}

\bibitem[{{Liu} {et~al.}(2024{\natexlab{a}}){Liu}, {Jiang}, {Feng}, {Zuo}, \& {Wang}}]{LiuJF2024MNRAS}
{Liu}, Q., {Jiang}, C., {Feng}, X., {Zuo}, P., \& {Wang}, Y. 2024{\natexlab{a}}, \mnras, 533, L25, \dodoi{10.1093/mnrasl/slae057}

\bibitem[{{Liu} {et~al.}(2015){Liu}, {Hu}, {Wang}, {Yang}, {Zhu}, {Liu}, {Luhmann}, \& {Richardson}}]{LiuHR2015ApJ}
{Liu}, Y.~D., {Hu}, H., {Wang}, R., {et~al.} 2015, \apjl, 809, L34, \dodoi{10.1088/2041-8205/809/2/L34}

\bibitem[{{Liu} {et~al.}(2024{\natexlab{b}}){Liu}, {Hu}, {Zhao}, {Chen}, \& {Wang}}]{LiuHZ2024ApJ}
{Liu}, Y.~D., {Hu}, H., {Zhao}, X., {Chen}, C., \& {Wang}, R. 2024{\natexlab{b}}, \apjl, 974, L8, \dodoi{10.3847/2041-8213/ad7ba4}

\bibitem[{{Liu} {et~al.}(2017){Liu}, {Hu}, {Zhu}, {Luhmann}, \& {Vourlidas}}]{LiuHZ2017ApJ}
{Liu}, Y.~D., {Hu}, H., {Zhu}, B., {Luhmann}, J.~G., \& {Vourlidas}, A. 2017, \apj, 834, 158, \dodoi{10.3847/1538-4357/834/2/158}

\bibitem[{{Liu} {et~al.}(2014){Liu}, {Luhmann}, {Kajdi{\v{c}}}, {Kilpua}, {Lugaz}, {Nitta}, {M{\"o}stl}, {Lavraud}, {Bale}, {Farrugia}, \& {Galvin}}]{LiuLK2014NatCo}
{Liu}, Y.~D., {Luhmann}, J.~G., {Kajdi{\v{c}}}, P., {et~al.} 2014, Nature Communications, 5, 3481, \dodoi{10.1038/ncomms4481}

\bibitem[{{Low}(1984)}]{Low1984ApJ}
{Low}, B.~C. 1984, \apj, 281, 392, \dodoi{10.1086/162110}

\bibitem[{{Lugaz} {et~al.}(2012){Lugaz}, {Farrugia}, {Davies}, {M{\"o}stl}, {Davis}, {Roussev}, \& {Temmer}}]{LugazFD2012ApJ}
{Lugaz}, N., {Farrugia}, C.~J., {Davies}, J.~A., {et~al.} 2012, \apj, 759, 68, \dodoi{10.1088/0004-637X/759/1/68}

\bibitem[{{Lugaz} {et~al.}(2017){Lugaz}, {Temmer}, {Wang}, \& {Farrugia}}]{LugazTW2017SoPh}
{Lugaz}, N., {Temmer}, M., {Wang}, Y., \& {Farrugia}, C.~J. 2017, \solphys, 292, 64, \dodoi{10.1007/s11207-017-1091-6}

\bibitem[{{Ma} {et~al.}(2011){Ma}, {Raymond}, {Golub}, {Lin}, {Chen}, {Grigis}, {Testa}, \& {Long}}]{MaRG2011ApJ}
{Ma}, S., {Raymond}, J.~C., {Golub}, L., {et~al.} 2011, \apj, 738, 160, \dodoi{10.1088/0004-637X/738/2/160}

\bibitem[{{MacQueen} {et~al.}(1986){MacQueen}, {Hundhausen}, \& {Conover}}]{MacQueenHC1986JGR}
{MacQueen}, R.~M., {Hundhausen}, A.~J., \& {Conover}, C.~W. 1986, \jgr, 91, 31, \dodoi{10.1029/JA091iA01p00031}

\bibitem[{{Majumdar} {et~al.}(2024){Majumdar}, {D'Huys}, {Mierla}, {Vashishtha}, {Talpeanu}, {Banerjee}, \& {Reiss}}]{MajumdarDM2024ApJ}
{Majumdar}, S., {D'Huys}, E., {Mierla}, M., {et~al.} 2024, \apjl, 970, L17, \dodoi{10.3847/2041-8213/ad5da5}

\bibitem[{{Mari{\v{c}}i{\'c}} {et~al.}(2009){Mari{\v{c}}i{\'c}}, {Vr{\v{s}}nak}, \& {Ro{\v{s}}a}}]{MaricicVR2009SoPh}
{Mari{\v{c}}i{\'c}}, D., {Vr{\v{s}}nak}, B., \& {Ro{\v{s}}a}, D. 2009, \solphys, 260, 177, \dodoi{10.1007/s11207-009-9421-y}

\bibitem[{{McCauley} {et~al.}(2015){McCauley}, {Su}, {Schanche}, {Evans}, {Su}, {McKillop}, \& {Reeves}}]{McCauleySS2015SoPh}
{McCauley}, P.~I., {Su}, Y.~N., {Schanche}, N., {et~al.} 2015, \solphys, 290, 1703, \dodoi{10.1007/s11207-015-0699-7}

\bibitem[{{Michalek} {et~al.}(2023){Michalek}, {Gopalswamy}, {Yashiro}, \& {Koleva}}]{MichalekGY2023ApJ}
{Michalek}, G., {Gopalswamy}, N., {Yashiro}, S., \& {Koleva}, K. 2023, \apj, 956, 59, \dodoi{10.3847/1538-4357/acf28d}

\bibitem[{{M{\"o}stl} {et~al.}(2015){M{\"o}stl}, {Rollett}, {Frahm}, {Liu}, {Long}, {Colaninno}, {Reiss}, {Temmer}, {Farrugia}, {Posner}, {Dumbovi{\'c}}, {Janvier}, {D{\'e}moulin}, {Boakes}, {Devos}, {Kraaikamp}, {Mays}, \& {Vr{\v{s}}nak}}]{MoestlRF2015NatCo}
{M{\"o}stl}, C., {Rollett}, T., {Frahm}, R.~A., {et~al.} 2015, Nature Communications, 6, 7135, \dodoi{10.1038/ncomms8135}

\bibitem[{{M{\"u}ller} {et~al.}(2017){M{\"u}ller}, {Nicula}, {Felix}, {Verstringe}, {Bourgoignie}, {Csillaghy}, {Berghmans}, {Jiggens}, {Garc{\'\i}a-Ortiz}, {Ireland}, {Zahniy}, \& {Fleck}}]{MuellerNF2017AA}
{M{\"u}ller}, D., {Nicula}, B., {Felix}, S., {et~al.} 2017, \aap, 606, A10, \dodoi{10.1051/0004-6361/201730893}

\bibitem[{{Myers} {et~al.}(2015){Myers}, {Yamada}, {Ji}, {Yoo}, {Fox}, {Jara-Almonte}, {Savcheva}, \& {Deluca}}]{MyersYJ2015Natur}
{Myers}, C.~E., {Yamada}, M., {Ji}, H., {et~al.} 2015, \nat, 528, 526, \dodoi{10.1038/nature16188}

\bibitem[{{Panasenco} {et~al.}(2013){Panasenco}, {Martin}, {Velli}, \& {Vourlidas}}]{PanasencoMV2013SoPh}
{Panasenco}, O., {Martin}, S.~F., {Velli}, M., \& {Vourlidas}, A. 2013, \solphys, 287, 391, \dodoi{10.1007/s11207-012-0194-3}

\bibitem[{{Patsourakos} \& {Vourlidas}(2009)}]{PatsourakosV2009ApJ}
{Patsourakos}, S., \& {Vourlidas}, A. 2009, \apjl, 700, L182, \dodoi{10.1088/0004-637X/700/2/L182}

\bibitem[{{Reva} {et~al.}(2024){Reva}, {Loboda}, {Bogachev}, \& {Kirichenko}}]{RevaLB2024SoPh}
{Reva}, A., {Loboda}, I., {Bogachev}, S., \& {Kirichenko}, A. 2024, \solphys, 299, 55, \dodoi{10.1007/s11207-024-02302-1}

\bibitem[{{Riley} {et~al.}(2018){Riley}, {Baker}, {Liu}, {Verronen}, {Singer}, \& {G{\"u}del}}]{RileyBL2018SSRv}
{Riley}, P., {Baker}, D., {Liu}, Y.~D., {et~al.} 2018, \ssr, 214, 21, \dodoi{10.1007/s11214-017-0456-3}

\bibitem[{{Sahade} {et~al.}(2025){Sahade}, {Vourlidas}, \& {Mac Cormack}}]{SahadeVM2025ApJ}
{Sahade}, A., {Vourlidas}, A., \& {Mac Cormack}, C. 2025, \apj, 978, 41, \dodoi{10.3847/1538-4357/ad96ba}

\bibitem[{{Savani} {et~al.}(2010){Savani}, {Owens}, {Rouillard}, {Forsyth}, \& {Davies}}]{SavaniOR2010ApJ}
{Savani}, N.~P., {Owens}, M.~J., {Rouillard}, A.~P., {Forsyth}, R.~J., \& {Davies}, J.~A. 2010, \apjl, 714, L128, \dodoi{10.1088/2041-8205/714/1/L128}

\bibitem[{{Scherrer} {et~al.}(2012){Scherrer}, {Schou}, {Bush}, {Kosovichev}, {Bogart}, {Hoeksema}, {Liu}, {Duvall}, {Zhao}, {Title}, {Schrijver}, {Tarbell}, \& {Tomczyk}}]{ScherrerSB2012SoPh}
{Scherrer}, P.~H., {Schou}, J., {Bush}, R.~I., {et~al.} 2012, \solphys, 275, 207, \dodoi{10.1007/s11207-011-9834-2}

\bibitem[{{Sheeley} {et~al.}(1999){Sheeley}, {Walters}, {Wang}, \& {Howard}}]{SheeleyWW1999JGR}
{Sheeley}, N.~R., {Walters}, J.~H., {Wang}, Y.~M., \& {Howard}, R.~A. 1999, \jgr, 104, 24739, \dodoi{10.1029/1999JA900308}

\bibitem[{{Shen} {et~al.}(2011){Shen}, {Wang}, {Gui}, {Ye}, \& {Wang}}]{ShenWG2011SoPh}
{Shen}, C., {Wang}, Y., {Gui}, B., {Ye}, P., \& {Wang}, S. 2011, \solphys, 269, 389, \dodoi{10.1007/s11207-011-9715-8}

\bibitem[{{Shen} {et~al.}(2012){Shen}, {Wu}, {Feng}, \& {Wu}}]{ShenWF2012JGRA}
{Shen}, F., {Wu}, S.~T., {Feng}, X., \& {Wu}, C.-C. 2012, Journal of Geophysical Research (Space Physics), 117, A11101, \dodoi{10.1029/2012JA017776}

\bibitem[{{Song} {et~al.}(2019){Song}, {Zhang}, {Cheng}, {Li}, {Tang}, {Wang}, {Zheng}, \& {Chen}}]{SongZC2019ApJ}
{Song}, H.~Q., {Zhang}, J., {Cheng}, X., {et~al.} 2019, \apj, 883, 43, \dodoi{10.3847/1538-4357/ab304c}

\bibitem[{{Sun} {et~al.}(2012){Sun}, {Hoeksema}, {Liu}, {Chen}, \& {Hayashi}}]{SunHL2012ApJ}
{Sun}, X., {Hoeksema}, J.~T., {Liu}, Y., {Chen}, Q., \& {Hayashi}, K. 2012, \apj, 757, 149, \dodoi{10.1088/0004-637X/757/2/149}

\bibitem[{{Thernisien} {et~al.}(2006){Thernisien}, {Howard}, \& {Vourlidas}}]{ThernisienHV2006ApJ}
{Thernisien}, A.~F.~R., {Howard}, R.~A., \& {Vourlidas}, A. 2006, \apj, 652, 763, \dodoi{10.1086/508254}

\bibitem[{{Thompson} {et~al.}(2012){Thompson}, {Kliem}, \& {T{\"o}r{\"o}k}}]{ThompsonKT2012SoPh}
{Thompson}, W.~T., {Kliem}, B., \& {T{\"o}r{\"o}k}, T. 2012, \solphys, 276, 241, \dodoi{10.1007/s11207-011-9868-5}

\bibitem[{{Verbeke} {et~al.}(2023){Verbeke}, {Mays}, {Kay}, {Riley}, {Palmerio}, {Dumbovi{\'c}}, {Mierla}, {Scolini}, {Temmer}, {Paouris}, {Balmaceda}, {Cremades}, \& {Hinterreiter}}]{VerbekeMK2023AdSpR}
{Verbeke}, C., {Mays}, M.~L., {Kay}, C., {et~al.} 2023, Advances in Space Research, 72, 5243, \dodoi{10.1016/j.asr.2022.08.056}

\bibitem[{{Vourlidas} {et~al.}(2011){Vourlidas}, {Colaninno}, {Nieves-Chinchilla}, \& {Stenborg}}]{VourlidasCN2011ApJ}
{Vourlidas}, A., {Colaninno}, R., {Nieves-Chinchilla}, T., \& {Stenborg}, G. 2011, \apjl, 733, L23, \dodoi{10.1088/2041-8205/733/2/L23}

\bibitem[{{Vourlidas} {et~al.}(2010){Vourlidas}, {Howard}, {Esfandiari}, {Patsourakos}, {Yashiro}, \& {Michalek}}]{VourlidasHE2010ApJ}
{Vourlidas}, A., {Howard}, R.~A., {Esfandiari}, E., {et~al.} 2010, \apj, 722, 1522, \dodoi{10.1088/0004-637X/722/2/1522}

\bibitem[{{Vourlidas} {et~al.}(2013){Vourlidas}, {Lynch}, {Howard}, \& {Li}}]{VourlidasLH2013SoPh}
{Vourlidas}, A., {Lynch}, B.~J., {Howard}, R.~A., \& {Li}, Y. 2013, \solphys, 284, 179, \dodoi{10.1007/s11207-012-0084-8}

\bibitem[{{Wang} {et~al.}(2018){Wang}, {Feng}, \& {Zhao}}]{WangFZ2018AA}
{Wang}, J., {Feng}, H., \& {Zhao}, G. 2018, \aap, 616, A41, \dodoi{10.1051/0004-6361/201731807}

\bibitem[{{Wang} {et~al.}(2015){Wang}, {Liu}, {Dai}, {Yang}, {Huang}, \& {Hu}}]{WangLD2015ApJ}
{Wang}, R., {Liu}, Y.~D., {Dai}, X., {et~al.} 2015, \apj, 814, 80, \dodoi{10.1088/0004-637X/814/1/80}

\bibitem[{{Wang} {et~al.}(2004){Wang}, {Shen}, {Wang}, \& {Ye}}]{WangSW2004SoPh}
{Wang}, Y., {Shen}, C., {Wang}, S., \& {Ye}, P. 2004, \solphys, 222, 329, \dodoi{10.1023/B:SOLA.0000043576.21942.aa}

\bibitem[{{Webb} \& {Howard}(2012)}]{WebbH2012LRSP}
{Webb}, D.~F., \& {Howard}, T.~A. 2012, Living Reviews in Solar Physics, 9, 3, \dodoi{10.12942/lrsp-2012-3}

\bibitem[{{Welsch}(2018)}]{Welsch2018SoPh}
{Welsch}, B.~T. 2018, \solphys, 293, 113, \dodoi{10.1007/s11207-018-1329-y}

\bibitem[{{Xing} {et~al.}(2024){Xing}, {Aulanier}, {Cheng}, {Xia}, \& {Ding}}]{XingAC2024ApJ}
{Xing}, C., {Aulanier}, G., {Cheng}, X., {Xia}, C., \& {Ding}, M. 2024, \apj, 966, 70, \dodoi{10.3847/1538-4357/ad2ea9}

\bibitem[{{Yan} {et~al.}(2018){Yan}, {Yang}, {Xue}, {Mei}, {Kong}, {Wang}, \& {Li}}]{YanYX2018ApJ}
{Yan}, X.~L., {Yang}, L.~H., {Xue}, Z.~K., {et~al.} 2018, \apjl, 853, L18, \dodoi{10.3847/2041-8213/aaa6c2}

\bibitem[{{Zhang} {et~al.}(2004){Zhang}, {Dere}, {Howard}, \& {Vourlidas}}]{ZhangDH2004ApJ}
{Zhang}, J., {Dere}, K.~P., {Howard}, R.~A., \& {Vourlidas}, A. 2004, \apj, 604, 420, \dodoi{10.1086/381725}

\bibitem[{{Zhang} {et~al.}(2024){Zhang}, {Ou}, {Huang}, {Song}, \& {Ma}}]{ZhangOH2024ApJ}
{Zhang}, Q., {Ou}, Y., {Huang}, Z., {Song}, Y., \& {Ma}, S. 2024, \apj, 977, 4, \dodoi{10.3847/1538-4357/ad8bad}

\bibitem[{{Zhang} {et~al.}(2022){Zhang}, {Chen}, {Li}, {Lu}, \& {Li}}]{ZhangCL2022SoPh}
{Zhang}, Q.~M., {Chen}, J.~L., {Li}, S.~T., {Lu}, L., \& {Li}, D. 2022, \solphys, 297, 18, \dodoi{10.1007/s11207-022-01952-3}

\bibitem[{{Zhao} {et~al.}(2019){Zhao}, {Liu}, {Hu}, \& {Wang}}]{ZhaoLH2019ApJ}
{Zhao}, X., {Liu}, Y.~D., {Hu}, H., \& {Wang}, R. 2019, \apj, 882, 122, \dodoi{10.3847/1538-4357/ab379b}

\bibitem[{{Zheng} {et~al.}(2023){Zheng}, {Liu}, {Liu}, {Wang}, {Hou}, {Feng}, {Kong}, {Huang}, {Song}, {Tian}, {Chen}, {Erd{\'e}lyi}, \& {Chen}}]{ZhengLL2023ApJ}
{Zheng}, R., {Liu}, Y., {Liu}, W., {et~al.} 2023, \apjl, 949, L8, \dodoi{10.3847/2041-8213/acd0ac}

\bibitem[{{Zhong} {et~al.}(2025){Zhong}, {Chen}, {Ni}, {Chen}, {Zheng}, {Kong}, \& {Li}}]{ZhongCN2025ApJ}
{Zhong}, Z., {Chen}, Y., {Ni}, Y.~W., {et~al.} 2025, \apj, 980, 42, \dodoi{10.3847/1538-4357/ada387}

\bibitem[{{Zhong} {et~al.}(2021){Zhong}, {Guo}, \& {Ding}}]{ZhongD2021NatCo}
{Zhong}, Z., {Guo}, Y., \& {Ding}, M.~D. 2021, Nature Communications, 12, 2734, \dodoi{10.1038/s41467-021-23037-8}

\bibitem[{{Zhu} {et~al.}(2020){Zhu}, {Qiu}, {Liewer}, {Vourlidas}, {Spiegel}, \& {Hu}}]{ZhuQL2020ApJ}
{Zhu}, C., {Qiu}, J., {Liewer}, P., {et~al.} 2020, \apj, 893, 141, \dodoi{10.3847/1538-4357/ab838a}

\bibitem[{{Zuccarello} {et~al.}(2012){Zuccarello}, {Bemporad}, {Jacobs}, {Mierla}, {Poedts}, \& {Zuccarello}}]{ZuccarelloBJ2012ApJ}
{Zuccarello}, F.~P., {Bemporad}, A., {Jacobs}, C., {et~al.} 2012, \apj, 744, 66, \dodoi{10.1088/0004-637X/744/1/66}

\end{thebibliography}
\end{document}